\documentclass[traditabstract]{aa}
\usepackage{epsfig}    
\usepackage{lscape}
\usepackage{natbib}
\bibpunct{(}{)}{;}{a}{}{,}    %% natbib format like A&A and ApJ
\usepackage{graphics}
\usepackage{graphicx,graphics}
\usepackage{color} 
\usepackage{times}
\usepackage{amsmath}
\usepackage{amssymb}
\begin{document}

\title {Properties of galaxies with an offset between the position angles of the major kinematic and photometric axes}
      
\author {
         L.~S.~Pilyugin\inst{\ref{MAO},\ref{ARI}} \and 
         E.~K.~Grebel\inst{\ref{ARI}} \and
         I.~A.~Zinchenko\inst{\ref{MAO},\ref{ARI}} \and
         J.~M.~V\'{i}lchez\inst{\ref{IAA}} \and
         F.~Sakhibov\inst{\ref{UAS}} \and
         Y.~A.~Nefedyev\inst{\ref{KGU}} \and
         P.~P.~Berczik\inst{\ref{MAO},\ref{ARI}}
         }
\institute{Main Astronomical Observatory, National Academy of Sciences of Ukraine, 27 Akademika Zabolotnoho St, 03680, Kiev, Ukraine \label{MAO} \and
Astronomisches Rechen-Institut, Zentrum f\"{u}r Astronomie der Universit\"{a}t Heidelberg, M\"{o}nchhofstr.\ 12--14, 69120 Heidelberg, Germany \label{ARI} \and
Instituto de Astrof\'{\i}sica de Andaluc\'{\i}a, CSIC, Apdo 3004, 18080 Granada, Spain \label{IAA} \and
University of Applied Sciences of Mittelhessen, Campus Friedberg,Department of Mathematics, Natural Sciences and Data Processing, 
Wilhelm-Leuschner-Stra\ss e 13, 61169 Friedberg, Germany \label{UAS} \and
Kazan Federal University, 18 Kremlyovskaya St., 420008, Kazan, Russian Federation \label{KGU} }

\abstract{ 
We derive the photometric, kinematic, and abundance characteristics of
18 star-forming MaNGA galaxies with fairly regular velocity fields and
surface brightness distributions and with a large offset between the
measured position angles of the major kinematic and photometric axes,
$\Delta$PA $\ga$ 20$\degr$. The aim is to examine if there is any other
distinctive characteristic common to these galaxies.  We found
morphological signs of interaction in some (in 11 out of 18) but not
in all galaxies. The observed velocity fields show a large variety;
the maps of the isovelocities vary from an hourglass-like appearance
to a set of straight lines.  The position angles of the major
kinematic axes of the stellar and gas rotations are close to each
other.  The values of the central oxygen abundance, radial abundance
gradient, and star formation rate are distributed within the intervals
defined by galaxies with small (no) $\Delta$PA of similar mass.  Thus,
we do not find any specific characteristic common to all galaxies with
large $\Delta$PA.  Instead, the properties of these galaxies are
similar to those of galaxies with small (no) $\Delta$PA.  This
suggests that either the reason responsible for the large $\Delta$PA
does not influence other characteristics or the galaxies with large
$\Delta$PA do not share a common origin, they can, instead, originate
through different channels.
}

\keywords{galaxies: kinematics and dynamics -- galaxies: abundances -- ISM}

\titlerunning{Galaxies with offset kinematic and photometric axes}
\authorrunning{Pilyugin et al.}
\maketitle

\section{Introduction}
%=====================

The position angle of the photometric major axis PA$_{phot}$ usually
coincides with (or at least, is close to) the position angle of the
kinematic major axis PA$_{kin}$ in disk galaxies.  \citet{Barnes2003}
estimated that non-axisymmetric features, such as spirals and bars,
introduce an average position angle uncertainty of $\sim$5$\degr$. A
large offset between the photometric and kinematic position angles is
believed to be caused by interactions or mergers of galaxies and,
consequently, to be an indicator of such events
\citep{Barrera2014,Barrera2015,Bloom2017,Rodrigues2017}.  The
condition that there is no mismatch between the kinematic and
photometric PAs is among the criteria that serve to properly
distinquish rotating discs from interacting or merging galaxies
\citep{Rodrigues2017}.

\citet{Garrido2005} and \citet{Epinat2008} derived the kinematic
position angles for a sample of spiral galaxies from Fabry-Perot
observations obtained in the framework of GHASP (Gassendi HAlpha
survey of SPirals).  They compared their kinematical PAs with the
photometric PAs from the
HyperLeda\footnote{http://leda.univ-lyon1.fr/} database and found that
the offset between the photometric and kinematic position angles,
$\Delta$PA, is observed in both isolated as well as in interacting
galaxies.  \citet{Barrera2014,Barrera2015} determined the offsets
between the photometric and kinematic position angles for isolated and
interacting or merging galaxies measured by the Calar Alto Legacy
Integral Field Area (CALIFA) survey
\citep{Sanchez2012,Husemann2013,GarciaBenito2015}. They found that
$\Delta$PA $> 21\degr$ for 43\% of the interacting or merging galaxies
and for 10\% of the isolated galaxies in their sample.
\citet{Graham2018} found that for 84.5\% of the MaNGA \citep[Mapping
Nearby Galaxies at Apache Point Observatory,][]{Bundy2015} spiral
galaxies the axes are aligned within 10$\degr$.

The presence of a large $\Delta$PA in some isolated galaxies means
that an interaction or merger is not the necessary condition for the
phenomenon of the large $\Delta$PA.  Moreover, the lack of a large
$\Delta$PA in a fraction of interacting galaxies means that
interactions are not a sufficient condition for the appearance of a
large $\Delta$PA in a galaxy, that is, it would seem that not every
interaction does result in a remnant with a large $\Delta$PA but that
only certain interactions may produce such a galaxy.

Morphology, kinematics, heavy element abundances, and other properties
are affected by interactions and mergers during galaxy evolution
\citep[e.g.,][]{Toomre1972,Veilleux2002,Rupke2010a,Rupke2010b,Rich2012,Larson2016}.
If a large offset between the photometric and kinematic position
angles is induced by interactions then one can expect that other signs
of the interaction can be found in those galaxies as well.  We examine
the photometric, kinematic, and abundance properties of 18 MaNGA
galaxies with a large offset between the position angles of the major
kinematic and photometric axes ($\Delta$PA $\ga$ 20$\degr$) in order
to ascertain whether their large $\Delta$PA is the only peculiar
property of those galaxies or whether other properties are also
unusual or at least similar for all these galaxies.  It should be
emphasized that only galaxies with regular velocity fields (surface
brightness distributions) are considered so that a unique major
kinematic (photometric) axis can be defined for the object as whole. 

The paper is organized in the following way. The data are described in
Section 2. In Section 3 the properties of the galaxies are discussed.
Section 4 contains a brief summary.

\section{Target galaxies}
%=====================

\subsection{Data}
%=====================

%====================================    Fig  No 01   Maps M-8454-12701 
\begin{figure*}
\resizebox{1.00\hsize}{!}{\includegraphics[angle=000]{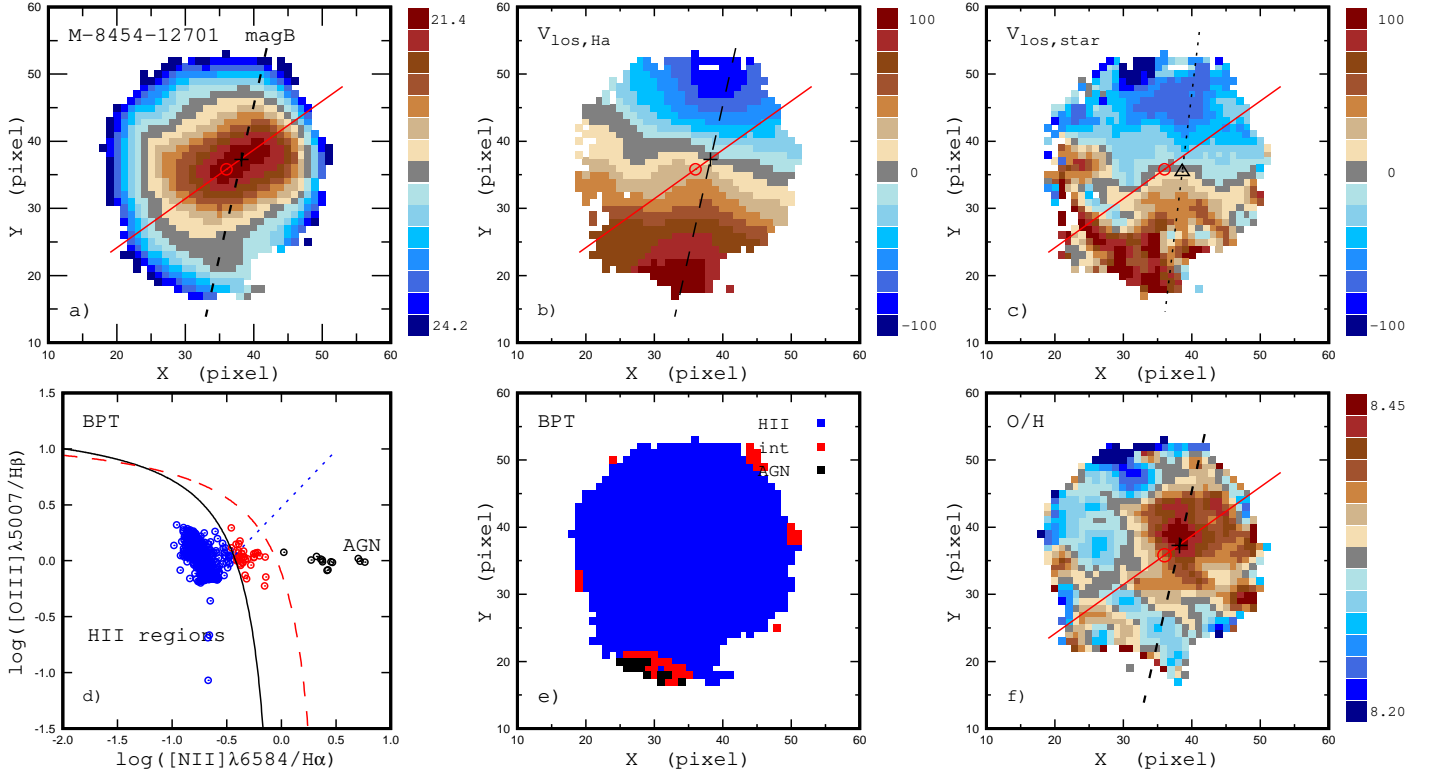}}
\caption{
Properties of the MaNGA galaxy M-8454-12701.
{\em Panel} $a$ shows the surface brightness distribution across the
image of the galaxy in sky coordinates (pixels). The value of the
surface brightness is color-coded. The circle shows the photometric
center of the galaxy, the solid line indicates the position of the
major photometric axis of the galaxy. 
{\em Panel} $b$ shows the observed (line-of-sight) H${\alpha}$ velocity field
in sky coordinates. The plus sign denotes the kinematic
center of the galaxy determined from the H${\alpha}$ velocity field.
The dashed line indicates the position of the major kinematic  (H${\alpha}$)
axis of the galaxy.
{\em Panel} $c$ shows the observed stellar velocity field.
The triangle denotes the kinematic
center of the galaxy determined from the stellar velocity field,
and the dotted line indicates the position of the kinematic  major
(stellar) axis of the galaxy.
{\em Panel} $d$ shows the BPT diagram. The symbols are individual spaxels.
Solid and long-dashed curves mark the demarcation line between AGNs and H\,{\sc ii}
regions defined by \citet{Kauffmann2003} and \citet{Kewley2001}, respectively.
The short-dashed line is the dividing line between Seyfert galaxies and
LINERs defined by \citet{CidFernandes2010}.
The black points are AGN-like objects according to the dividing line of  \citet{Kewley2001}.
The blue points are the  H\,{\sc ii}-region-like objects according to the
dividing line of  \citet{Kauffmann2003}.
The red points are intermediate objects located between the dividing lines
of  \citet{Kauffmann2003} and \citet{Kewley2001}.
{\em Panel} $e$ shows the locations of the spaxels with the AGN-like,  H\,{\sc ii}-region-like,
and intermediate spectra on the image of the galaxy.
{\em Panel} $f$ shows the abundance distribution  across the
image of the galaxy in sky coordinates (pixels).
}
\label{figure:m-8454-12701}
\end{figure*}

The spectroscopic measurements from the SDSS-IV MaNGA survey
\citep{Bundy2015,Albareti2017} provide the possibility to measure the
surface brightness distribution and to determine optical radii and
luminosities, to measure the observed gas and stellar velocity fields
and to derive kinematic angles, and to measure emission line fluxes
and obtain abundance maps. In our previous paper \citep{Pilyugin2019},
we derived rotation curves, surface brightness profiles, and oxygen
abundance distributions for star-forming galaxies using the publicly
available spectroscopy obtained by the SDSS-IV MaNGA survey.  A sample
of 147 galaxies with offsets between the position angles of the
kinematic and photometric axes $\Delta$PA less than $\sim$20$\degr$
were considered to examine the relations between the abundance
properties, rotation velocity, and other macroscopic properties, for example,
stellar mass \citep{Pilyugin2019}.

Here we will investigate galaxies with a large offset between the
position angles of the kinematic and photometric major axes.  We
selected a sample of MaNGA galaxies to be analyzed by considering
their derived gas velocity fields, surface brightnesses, and abundance
maps.  We selected those galaxies using the following criteria. \\

  They had to be star-forming galaxies since the characteristics 
  to be analyzed (gas velocities, abundances, star formation rate)
  are based on emission lines.  \\

  Galaxies measured with a small number of fibers (19 and 37) were excluded
  from consideration. \\

  We required that the surface brightness distribution and the gas 
  velocity field are rather regular in order be able to obtain reliable values 
  for the position angles of the kinematic and photometric major axes and to be 
  able to establish galactocentric distances for the spaxels in the galaxy 
  image.
  We also required that the spaxels with measured emission lines and surface
  brightness need to be distributed across the galactic disks.
  Under those conditions, the obtained characteristics (e.g., position angles,
  abundance distribution) can be interpreted as global properties, that is, 
  they can be used to characterize the galaxy as a whole.    
  The galaxy M-8551-09102 was measured up to a radius of 0.70 $R_{25}$, 
  the galaxy M-8454-12701 up to 0.77 $R_{25}$, and the spaxels with 
  measured emission lines cover more than $\sim$0.8 $R_{25}$ in the other 
  selected galaxies. 
  It should be stressed that only those spaxel spectra where all the used lines 
  ([O\,{\sc ii}]$\lambda$3727+$\lambda$3729, H$\beta$, 
  [O\,{\sc iii}]$\lambda$5007, H$\alpha$, [N\,{\sc ii}]$\lambda$6584)  
  were measured with a signal-to-noise ratio S/N $> 3$ were considered.  
  Therefore, the spaxels with reliably measured spectra may cover less of 
  the radial extent than the radius across which spaxel spectra are 
  available. \\

  We adopted the value of the offset between the position angles of the 
  kinematic and photometric major axes $\Delta$PA =  $\sim20\degr$ as 
  demarcation value between galaxies with small and large $\Delta$PA.
  Of course, the choice of the demarcation value is somewhat arbitrary.

Our final list includes 18 galaxies, which are listed in Table 
\ref{table:sample}.

\citet{Barrera2015} found that morpho-kinematic misalignments are
related to a particular stage of the merger event (pre-merger, merger,
post-merger, remnant).  The majority of galaxies with $\Delta$PA $>
20\degr$ in their sample are classified as galaxies in the merger
stage.  The median $\Delta$PA is maximum for those galaxies.  The
velocity field (and the surface brightness distribution) in a
galaxy in the merger stage is usually complex and a unique kinematic
(photometric) major axis cannot be defined for the object as whole.
Such galaxies do not meet our selection criteria.  Some of the
galaxies of \citet{Barrera2015} with a large $\Delta$PA show rather
regular velocity fields and surface brightness distributions (for example,
NGC~6977 and NGC~7738). Those galaxies are classified as galaxies in
pre-merger, post-merger, and remnant stages of the merger event
\citep{Barrera2015}.  However, the number of galaxies with large
misalignments in those merger stages is low; the median misalignments
are below or comparable to the demarcation value adopted here.  Only a
few interacting galaxies of the sample from \citet{Barrera2015} meet
our selection criteria.

%++++++++++++++++++ Table     Sample 
\setcounter{table}{0}
\begin{table*}
%\tabletypesize{\small}
%\tabletypesize{\scriptsize}
\caption[]{\label{table:sample}
Properties of our sample of MaNGA galaxies.  The columns show
the name (the MaNGA number),
the right ascension (RA) and declination (Dec) (J2000.0),
the galaxy distance $d$ in Mpc,
the spectroscopic stellar mass $M_{sp}$ in solar masses,
the optical radius $R_{25}$ in kpc,
the photometric inclination angle $i_{phot}$,
the photometric position angle of the major axis $PA_{phot}$, 
the kinematic (gas) position angle of the major axis $PA_{kin,H\alpha}$, 
the kinematic (star) position angle of the major axis $PA_{kin,star}$, 
the central oxygen abundance (O/H)$_{0}$,
and the radial oxygen abundance gradient in dex/$R_{25}$. 
}
\begin{center}
\begin{tabular}{lccccccccccc} \hline \hline
Name$^{a}$                   &
RA                    &
Dec                   &
$d$                   &
log$M_{sp}$           &
$R_{25}$              &
$i_{phot}$            &           
PA$_{phot}$           &           
PA$_{kin,H{\alpha}}$    &           
PA$_{kin,star}$        &           
(O/H)$_{0}^{b}$       &
grad(O/H)           \\
                      &
[$\degr$]            &           
[$\degr$]            &           
[Mpc]                  &
[$M_{\sun}$]             &
[kpc]                  &
[$\degr$]            &           
[$\degr$]            &           
[$\degr$]            &           
[$\degr$]            &           
                     &           
[dex/$R_{25}$]        \\   \hline
 8140 09101     &  116.507218 &   40.875058 &  157.9 &    9.61 &   8.04 &  40.9 &  368.6 &  339.2 &    1.2 &   8.45 &  -0.2995 \\
 8145 06103     &  116.503038 &   28.699633 &   98.7 &   10.03 &   9.09 &  43.8 &  260.9 &  345.7 &  343.2 &   8.66 &  -0.2177 \\
 8146 09102     &  117.811568 &   28.153630 &  215.6 &   10.79 &  12.54 &  55.1 &  349.2 &  381.6 &  379.4 &   8.68 &  -0.2074 \\
 8249 06101     &  137.562456 &   46.293269 &  113.3 &   10.25 &  10.44 &  63.8 &  286.3 &  239.4 &  243.5 &   8.67 &  -0.1594 \\
 8252 12704     &  145.943835 &   48.290713 &  106.2 &    9.77 &  13.13 &  51.6 &   96.0 &  178.0 &  178.5 &   8.59 &  -0.1809 \\
 8254 09102$^*$ &  163.701861 &   44.850180 &  159.6 &    9.70 &   7.74 &  51.7 &  136.9 &   79.0 &   73.3 &   8.56 &  -0.1524 \\
 8254 12703$^*$ &  162.051276 &   45.595289 &  210.0 &   10.12 &  12.73 &  57.5 &  174.4 &  206.8 &  211.8 &   8.52 &  -0.1724 \\
 8257 12701$^*$ &  165.495818 &   45.228024 &   87.6 &   10.20 &  12.53 &  56.6 &  246.0 &  272.3 &  277.2 &   8.64 &  -0.0647 \\
 8320 06104$^*$ &  206.554380 &   23.084053 &  131.1 &    9.42 &  11.12 &  50.2 &  155.7 &  182.2 &  175.0 &   8.41 &  -0.2301 \\
 8329 06104$^*$ &  213.110482 &   45.690410 &  121.9 &   10.66 &  10.34 &  41.7 &  127.0 &   92.0 &   93.3 &   8.64 &  -0.1659 \\
 8450 06102$^*$ &  171.748834 &   21.141675 &  174.8 &   10.30 &  13.56 &  43.1 &  297.5 &  352.6 &  349.0 &   8.60 &  -0.2249 \\
 8451 12703$^*$ &  164.028900 &   43.156568 &  160.4 &    9.50 &  12.44 &  47.9 &  211.7 &  179.3 &  173.1 &   8.42 &  -0.2503 \\
 8454 12701$^*$ &  154.005710 &   44.176240 &  187.0 &    9.87 &  11.79 &  27.8 &  126.1 &  167.5 &  173.2 &   8.42 &  -0.1761 \\
 8483 06103$^*$ &  247.179730 &   49.061080 &  151.0 &    9.19 &   6.95 &  39.0 &  308.1 &  343.0 &  334.2 &   8.30 &  -0.1242 \\
 8486 12702$^*$ &  235.915585 &   47.667443 &  161.4 &    9.60 &   7.82 &  30.8 &  228.5 &  250.5 &  261.4 &   8.50 &  -0.2598 \\
 8547 06102     &  217.324860 &   52.665559 &  131.0 &   10.53 &  14.92 &  60.5 &  339.1 &  315.5 &  317.4 &   8.69 &  -0.1082 \\
 8551 09102     &  235.669362 &   45.432085 &  160.6 &    9.40 &   8.95 &  61.2 &   90.8 &   59.7 &   53.7 &   8.47 &  -0.1416 \\
 8568 12703$^*$ &  156.810365 &   38.204502 &  221.2 &   10.47 &  16.09 &  26.8 &   77.9 &   41.3 &   34.3 &   8.68 &  -0.3692 \\
\hline
\end{tabular}\\
\end{center}
\begin{flushleft}
$^a$ galaxies with signs of an interaction are labeled by an asterisk \\
$^b$ as 12+log(O/H)
\end{flushleft}
\end{table*}

The SDSS data base provides values of the stellar masses of its target
galaxies determined in different ways.  We have chosen the
spectroscopic $M_{sp}$ masses of the SDSS and BOSS galaxies
\citep[BOSS stands for the Baryon Oscillation Spectroscopic Survey in
SDSS-III, see][]{Dawson2013}.  The spectroscopic masses, $M_{sp}$,
are the median (50th percentile of the probability distribution
function, PDF) of the logarithmic stellar masses from table {\sc
stellarMassPCAWiscBC03} determined by the Wisconsin method
\citep{Chen2012} with the stellar population synthesis models from
\citet{Bruzual2003}.

\subsection{Mapping the properties of our galaxies}
%=====================

The spectrum of each spaxel is reduced in the manner described in
\citet{Zinchenko2016}.  Briefly, the stellar background in all
spaxels is fitted using the public version of the STARLIGHT code
\citep{CidFernandes2005,Mateus2006,Asari2007} adapted for execution in
the NorduGrid ARC\footnote{http://www.nordugrid.org/} environment of
the Ukrainian National Grid.  To fit the stellar spectra we used 150
synthetic simple stellar population (SSP) spectra from the
evolutionary synthesis models by \citet{Bruzual2003} with ages from
1~Myr up to 13~Gyr and metallicities of $Z = 0.0001$, 0.004, 0.008,
0.02, and 0.05. We adopted the reddening law of \citet{Cardelli1989}
with $R_V = 3.1$.  The resulting stellar radiation contribution was
subtracted from the observed spectrum in order to measure and analyze
the line emission from the gaseous component. The line intensities
were measured using single Gaussian line profile fits on the pure
emission spectra. 

The total [O\,{\sc iii}]$\lambda$$\lambda$4959,5007 flux was 
estimated as $1.33 \: \times$~[O\,{\sc iii}]$\lambda$5007 instead of
the sum of the fluxes of both lines.  These lines originate from
transitions from the same energy level, so their flux ratio can be
determined by the transition probability ratio, which is very close to
3 \citep{Storey2000}.  The strongest line, [O\,{\sc
iii}]$\lambda$5007, can be measured with higher precision than the
weakest one. This is particularly important for high-metallicity
H\,{\sc ii} regions, which have weak high-excitation emission lines.
Similarly, the [N\,{\sc ii}]$\lambda$$\lambda$6548,6584 lines also
originate from transitions from the same energy level and the
transition probability ratio for those lines is again close to 3
\citep{Storey2000}. Therefore, we estimated its total flux as
$1.33$~[N\,{\sc ii}]$\lambda$6584. For each spectrum, we measure the
fluxes of the 
[O\,{\sc ii}]$\lambda\,\lambda$3727,3729,
H$\beta$,  
[O\,{\sc iii}]$\lambda$5007,
H$\alpha$,  
[N\,{\sc ii}]$\lambda$6584, and
[S\,{\sc ii}]$\lambda$6717, 6731 lines.
The emission line fluxes were corrected for interstellar reddening
using the theoretical H$\alpha$/H$\beta$ ratio and the reddening
function from \citet{Cardelli1989} for $R_{V}$ = 3.1.  We assume
$C_{{\rm H}{\beta}} = 0.47A_{V}$ \citep{Lee2005}. 

The surface brightness in the SDSS $g$ and $r$ bands for each spaxel
was obtained from broadband SDSS images created from the data cube.
The measured magnitudes were converted to $B$-band magnitudes and were
corrected for Galactic foreground extinction using the recalibrated
$A_V$ values of \citet{Schlafly2011} reported in the NASA/IPAC
Extragalactic Database ({\sc ned})\footnote{The NASA/IPAC
Extragalactic Database ({\sc ned}) is operated by the Jet Populsion
Laboratory, California Institute of Technology, under contract with
the National Aeronautics and Space Administration.  {\tt
http://ned.ipac.caltech.edu/}}.

The deprojected galaxy plane was divided into rings with a width
of one pixel.  The position angle of the major photometric axis and
the galaxy inclination were assumed to be the same for all the rings,
that is, constant within the disc.  The pixel coordinates of the
photometric center of the galaxy, the inclination angle $i$, the
position angle of the photometric major axis, $PA_{phot}$, and the
observed surface-brightness profile were derived through the best fit
to the measured surface-brightness map.  The observed
surface-brightness profile within a galaxy was fitted by a broken
exponential profile for the disc and by a general S\'{e}rsic profile
for the bulge \citep{Pilyugin2018}. The optical radius of the galaxy
$R_{25}$ was estimated using the obtained fit.    

Panel a of Fig.~\ref{figure:m-8454-12701} shows the obtained surface
brightness distribution across the image of the MaNGA galaxy
M-8454-12702 in sky coordinates (pixels). North is up and East to the
left.  The scale of a pixel is 0.5 arcsec.  The value of the surface
brightness is color-coded. The circle shows the photometric center of
the galaxy, and the solid line indicates the position of the major
photometric axis of the galaxy.  The obtained surface brightness
distributions across the images of the other galaxies of our sample
and the positions of the major photometric axes are shown in
Fig.~\ref{figure:m-8140-09101} -- Fig.~\ref{figure:m-8568-12703} in
the Appendix.  

\begin{figure}
\centering
\resizebox{1.00\hsize}{!}{\includegraphics[angle=000]{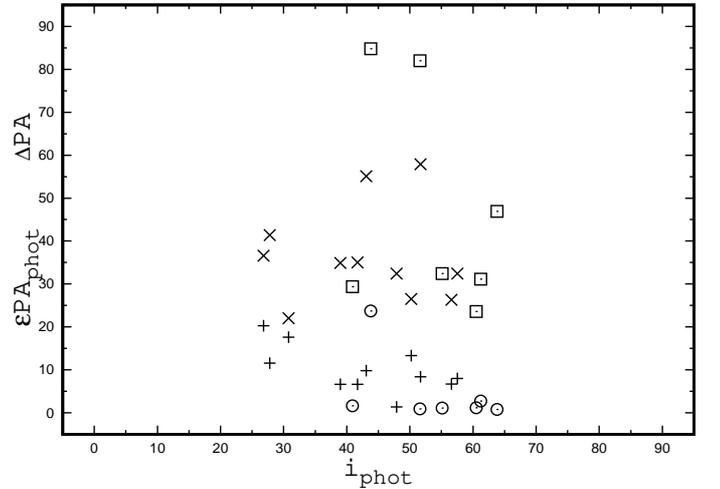}}
\caption{
Difference between the position angles of major photometric axis
obtained here and those listed in the NASA-Sloan Atlas,
$\epsilon$PA$_{phot}$, as a function of the inclination angle of the
galaxy for galaxies with (sign plus) and without (circles) signs of
interactions.  The difference between the position angles of the major
photometric and kinematic axes $\Delta$PA obtained here as a function
of the inclination angle for galaxies with (crosses) and without
(squares) signs of the interaction is also plotted.  The absolute
values of the differences are shown.
}
\label{figure:epa}
\end{figure}

The position angles of the phometric major axes of the MaNGA
galaxies based on SDSS photometry are given in the NASA-Sloan-Atlas
(NSA)\footnote{\tt
https://www.sdss.org/dr13/manga/manga-target-selection/nsa/}.  To
check the validity of the position angles derived here from the MaNGA
measurements we compared our values of the position angle with those
from the NSA.  We took the position angle of the major photometric
axis from the NSA S\'ersic model structural parameters, which were fit
using the SDSS r-band image.  Fig.~\ref{figure:epa} shows the absolute
value of the difference between the position angle obtained here and
that from the NSA, $\epsilon$PA, as a function of the inclination
angle of the galaxy.  We divide our galaxies into two groups, with and
without signs of interaction (see below).  The $\epsilon$PA for the
galaxies with interaction signatures are shown by plus signs and those
for seemingly non-interacting galaxies are denoted by circles.
Inspection of Fig.~\ref{figure:epa} shows that our PA$_{phot}$ for
galaxies without signs of interaction agree with the
SDSS-photometry-based values within $\sim3\degr$ with the one
exception of the galaxy M-8145-06103, which is a galaxy with bright
spiral arms.  The $\epsilon$PA for apparently interacting galaxies are
larger, within $\sim10\degr$ for galaxies with inclination angles $i
\ga 40\degr$ and up to $\sim20\degr$ for some galaxies with the
inclination angles of $i\la 30\degr$.

%====================================    Fig  No 2   dPAphot,kin vs ePA 

The measurement of the emission lines provides the gas velocity of
each region (spaxel).  Panel $b$ of Fig.~\ref{figure:m-8454-12701}
shows the color-coded observed (line-of-sight) H${\alpha}$ velocity
field in the galaxy M-8454-12702 in sky coordinates.  The pixel
coordinates of the kinematic center of the galaxy, the inclination
angle $i_{kin}$, and the position angle of the kinematic major axis,
$PA_{kin}$ are derived through the best fit to the measured velocity
field using the standard relation between the observed line-of-sight
velocities recorded on a set of pixel coordinates and the kinematic
parameters
\citep[e.g.,][]{Warner1973,Begeman1989,deBlok2008,Oh2018,Pilyugin2019}.
Again the deprojected galaxy plane is divided into annuli with a width
of one pixel.  The position angle of the major kinematic axis and the
galaxy inclination are assumed to be the same for all the rings. The
plus symbol in panel $b$ of Fig.~\ref{figure:m-8454-12701} shows the
gas (H${\alpha}$) kinematic center of the galaxy; the dashed line
indicates the position of the major gas kinematic axis of the galaxy
M-8454-12702.

Accurate rotation curves cannot be determined for all of our galaxies
because the curves of isovelocities in the measured velocity fields in
some galaxies resemble more a set of the straight lines than a set of
parabola-like curves (the hourglass-like picture for the rotation
disk). That prevents the determination of a reliable value of the
kinematic inclination angle of a galaxy.   Thus, only the position
angles of the gas and stellar kinematic major axis will be considered. 

Fig.~\ref{figure:epa} shows the absolute value of the difference
between the position angles of the major kinematic and photometric
axes, $\Delta$PA, as a function of the inclination angle of a galaxy.
The $\Delta$PA for galaxies with signs of an interaction are indicated
by crosses and for galaxies without apparent interactions by squares.
Inspection of Fig.~\ref{figure:epa} shows that the $\Delta$PA exceeds
significantly the $\epsilon$PA for our galaxies with the exception of
the galaxy M-8486-12702, which has a small inclination angle.  

The fitting of the stellar continuum allows us to estimate the stellar
velocity of each spaxel.  Panel $c$ of Fig.~\ref{figure:m-8454-12701}
shows the color-coded observed (line-of-sight) stellar velocity field
in the galaxy M-8454-12702 in sky coordinates. The triangle marks the
stellar kinematic center of the galaxy. The dotted line indicates the
position of the major stellar kinematic axis of the galaxy.

The standard [N\,{\sc ii}]$\lambda$6584/H$\alpha$ versus [O\,{\sc
iii}]$\lambda$5007/H$\beta$ diagram (the BPT classification diagram)
suggested by \citet{Baldwin1981} is used to separate different types
of emission-line objects according to their main excitation mechanism
(that is, starburst or AGN).  Panel $d$ of Fig.~\ref{figure:m-8454-12701}
shows the BPT diagram for the spaxels in the galaxy M-8454-12702.  The
solid curve is the dividing line suggested by \citet{Kauffmann2003},
and the long-dashed curve line is that suggested by
\citet{Kewley2001}.  The objects located in the BPT diagram left of
(below) the separation curve of \citet{Kauffmann2003} (blue points)
are referred to as objects with H\,{\sc ii}-region-like spectra, the
objects located in the BPT diagram right from (above) the separation
curve of \citet{Kewley2001} (black points) are referred to as objects
with AGN-like spectra, and the objects located in the BPT diagram
between those separation curves (red points) are called objects with
intermediate spectra.  The short-dashed line is the dividing line
between Seyfert galaxies and LINERs defined by
\citet{CidFernandes2010}.

Panel $e$ of Fig.~\ref{figure:m-8454-12701} shows the locations of
spaxels with AGN-like (black points), H\,{\sc ii}-region-like (blue
points), and intermediate (red points) spectra in the image of the
galaxy M-8454-12702.

It has been found that the three-dimensional $R$ calibration from
\citet{Pilyugin2016} produces reliable abundances for MaNGA galaxies,
that is, the  $R$-calibration produces abundances compatible to the 
$T_{e}$-based abundance scale and is workable 
  over the whole metallicity scale of H\,{\sc ii} regions   
\citep{Pilyugin2018,Pilyugin2019,Zinchenko2019a,Zinchenko2019b}.  We
use those relations for abundance determinations also here.  Panel $f$
of Fig.~\ref{figure:m-8454-12701} shows the oxygen abundance
distribution across the image of the galaxy M-8454-12702 in sky
coordinates (pixels). The value of the oxygen abundance is
color-coded.

\section{Properties of our sample of MaNGA galaxies}
%=====================

\subsection{Morphological signs of an interaction or merger}
%=====================

%====================================    Fig  No 3     SB profiles of galaxies 
\begin{figure*}
\centering
\resizebox{1.00\hsize}{!}{\includegraphics[angle=000]{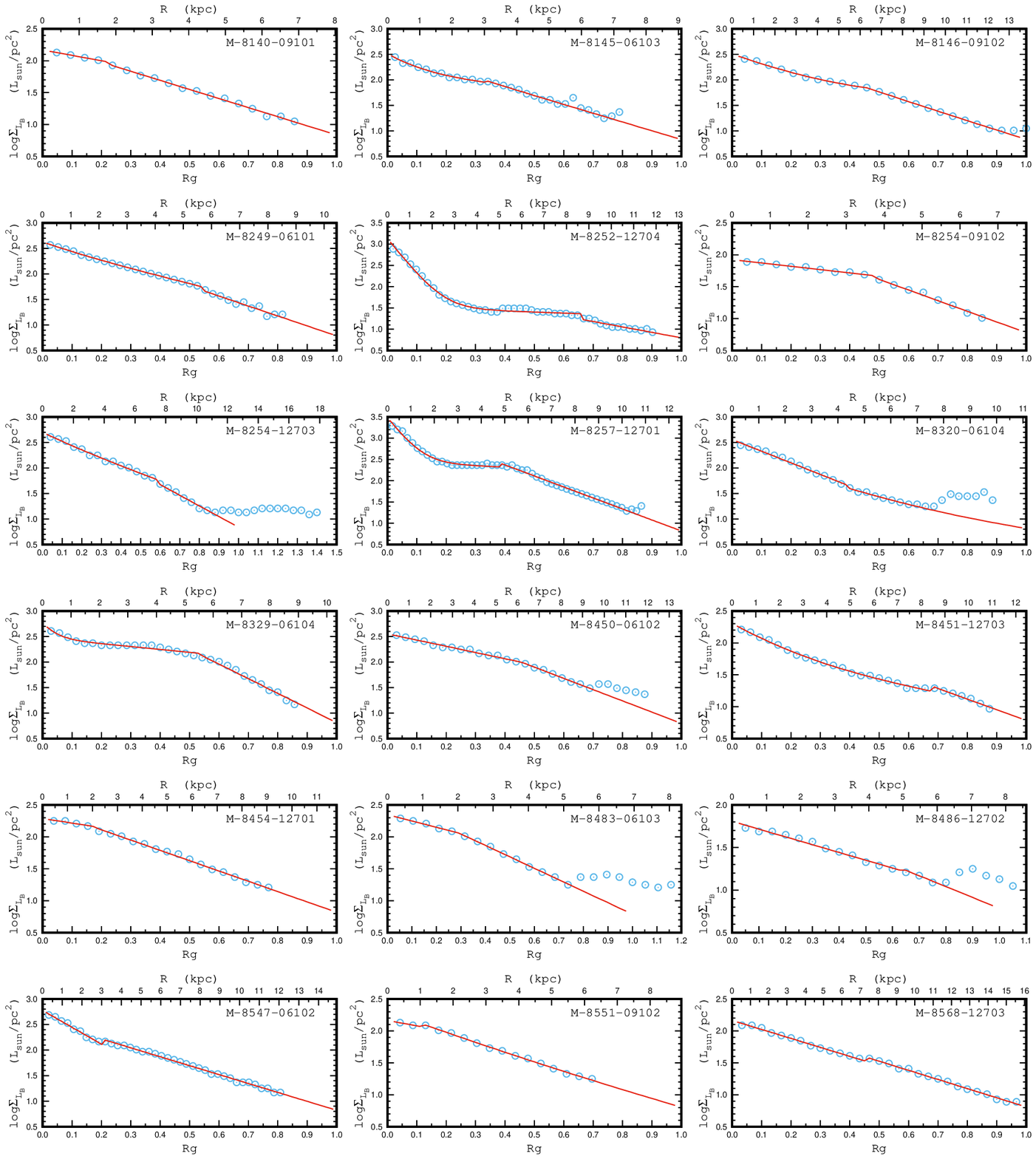}}
\caption{
Surface brightness profiles for our sample of MaNGA galaxies.
The points denote the observed profile. The line indicates the fit
by a broken exponential profile for the disc and by a
general S\'{e}rsic profile for the bulge. The lower OX axis marks
the fractional galactocentric distance normalized to the optical
radius, $R_{g} = R/R_{25}$. The upper OX axis shows 
the galactocentric distance expressed in kiloparsecs.
}
\label{figure:sb-profiles}
\end{figure*}

The morphologies of galaxies are widely used to recognize and classify
interacting or merging galaxies and the merger stages
\citep[e.g.,][]{Toomre1972,Veilleux2002,Barrera2015,Larson2016}.
Morphological classification schemes for the merger stages are based
on visual appearance.  The membership in a close galaxy pair and/or
the presence of tails, bridges, or other distortions are considered as
signs of an interaction or merger \citep[e.g.,][]{Martinez2010,
Morales2018}.  Those signatures can be considered as external
morphological signs of an interaction or merger.  Two nuclei and/or a
disturbed surface brightness distribution (merger-induced fine
structure such as ripples, box-like isophotes, or warped surface
brightness features \citep{Schweizer1992}) may be considered as internal
morphological signs of an interaction or merger. It should be noted that
there can be ambiguities in distinguishing between close galaxy pairs
and a galaxy with two nuclei due to projection effects. 

We examine the images of galaxies as represented by their surface
brightness distribution maps in Fig.~\ref{figure:m-8454-12701} and
Fig.~\ref{figure:m-8140-09101} -- Fig.~\ref{figure:m-8568-12703}, and
their surface brightness profiles in Fig.~\ref{figure:sb-profiles} in
order to search for morphological signatures of interactions or
mergers in our galaxies.

The surface brightness distribution within a galaxy is fitted by a
broken exponential profile for the disc and by a general S\'{e}rsic
profile for the bulge following to our previous paper
\citep{Pilyugin2018}.  We use the surface-brightness profile in solar
units for the bulge-disc decomposition. The magnitude of the Sun in
the $B$ band of the Vega photometric system,  $B_{\sun}$ = 5.45, was
taken from \citet{Blanton2007}.  The distances were adopted from {\sc
ned}.  The {\sc ned} distances use flow corrections for Virgo, the
Great Attractor, and Shapley Supercluster infall (adopting a
cosmological model with $H_{0} = 73$ km/s/Mpc, $\Omega_{m} = 0.27$,
and $\Omega_{\Lambda} = 0.73$).  Fig.~\ref{figure:sb-profiles} shows
the surface brightness profiles for our sample of the MaNGA galaxies.
The points denote the observed profile. The line indicates the fit by
a broken exponential profile for the disc and by a general S\'{e}rsic
profile for the bulge. The lower OX axis marks the fractional
galactocentric distance normalized to the optical radius, $R_{g} =
R/R_{25}$. The upper OX axis shows the galactocentric distance
expressed in kiloparsecs.

Below, we list morphological signs of an interaction or merger in
our selected galaxies (or the absence of such signatures).

{\em M-8140-09101.}
There are no obvious morphological signs of an interaction in this
galaxy.  However, a faint filamentary structure can be seen in the
north-west direction at a projected distance of approximately two
diameters of the galaxy in the sky plane.  The redshift of this
structure is not available, which prevents us from estimating the real
separation.

{\em M-8145-06103.}
M-8145-06103 (= CGCG 148-011) is an {\em Sbc} galaxy.
There are no obvious morphological signs of an interaction. 

{\em M-8146-09102.}
M-8146-09102 is an isolated galaxy \citep{ArgudoFernandez2015} without 
obvious morphological signs of an interaction.

{\em M-8249-06101.}
M-8249-06101 (= KUG~096+464) is an {\em SBb} galaxy.
There are no obvious morphological signs of an interaction. 

{\em M-8252-12704.}
M-8252-12704 (= KUG~0940+485) is an {\em Sb} galaxy.
There are no obvious morphological signs of an interaction. 

{\em M-8254-09102.}
The isophotes of M-8254-09102 are box-like, which is a morphological 
signature of an interaction. 

{\em M-8254-12703.}
There is a tail in the image of the galaxy, which indicates an
interaction. 

{\em M-8257-12701.}
M-8257-12701 (=UGC~6103 = Mrk~161) is an {\em SBc} galaxy.  The spot
(a second nucleus) in the surface brightness distribution (as well as
in the H$\alpha$ and stellar velocity fields, and in the abundance
distribution) can be seen in the image of M-8257-12701 towards the
north from the center.  This may be the projection of a close
companion (the difference between the values of the line-of sight
velocity of the center of the galaxy and the spot is around 70 km/s,
that is, is comparable to the variation of values of the
line-of-sight velocity in M-8257-12701 due to its rotation) or it
could be a second nucleus if the companion was already captured and is
now located within the main galaxy.  In any case, this is a
morphological sign of an interaction. 
  
{\em M-8320-06104.}
M-8320-06104 (= Mrk 795) is an {\em Sc} galaxy.
The optical image of M-8320-06104 is noticeably disturbed,
which is a morphological indication of an interaction. 

{\em M-8329-06104.}
M-8329-06104 (= PGC~50739) is an {\em SBbc} galaxy and a member of the
close galaxy pair UGC~9098 (PGC~50739 + PGC~50738).  Thus this galaxy
is likely interacting.

{\em M-8450-06102.}
M-8450-06102 (=CGCG~126-052) is an {\em Scd} galaxy.
There is a rise in the surface profile in the outer part of the galaxy. 
This may be interpreted as a sign of an interaction. 

{\em M-8451-12703.}
M-8451-12703 (= KUG~1053+434) is an {\em SBc} galaxy.
The isophotes are box-like, which is a morphological sign of an interaction. 

{\em M-8454-12701.}
The {\em Scd} galaxy M-8454-12701 (= KUG~1012+444) is a member of a pair. 
The components of the pair are very close to each other on the sky plane,
they almost overlap.  Hence this galaxy can be classified as an likely
interacting one.

{\em M-8483-06103.}
There is a tail, which is a morphological sign of an interaction. 

{\em M-8486-12702.}
The inner isophotes are box-like. There is a rise in the surface
brightness profile in the outer parts of the galaxy. Those are the
morphological indications of an interaction. 

{\em M-8547-06102.}
There are no obvious morphological signs of an interaction. 

{\em M-8551-09102.}
There are no obvious morphological signs of an interaction. 

{\em M-8568-12703.}
The galaxy  M-8568-12703  is a member of a galaxy pair.  The
components of the pair are very close to each other on the sky plane,
they overlap.  The north-east part of the image (X $<$ 40 and Y $>$ 50
pixel) is not included in the determinations of the kinematic and
photometric axes.  This galaxy can be classified as interacting one.

Thus, morphological signatures of an interaction are visible in 11 out
of 18   galaxies of our sample.  There are no obvious morphological
signs of an interaction in the remaining 7 galaxies of our sample, and
those galaxies are not members of close galaxy pairs.  It should be
noted that \citet{HernandezJimenez2015} found a misaglignment of
58$\degr$ between the photometric and kinematic axes in the main
galaxy of a pair where the main component is 20 times more luminous
than the secondary.

Simulations of interactions and mergers of galaxies were carried out
in many works \citep[][among many
others]{Walker1996,Naab2003,Bournaud2004,Springel2005,
Robertson2006,Governato2007,Lotz2008,Hopkins2009a,Hopkins2009b,
Zinchenko2015,RodriguezGomez2017}.  It is established that the
observed properties of the interaction or merger remnants depend on the
characteristics of progenitors, the geometry of the collision, and the
merger stage. The properties of the merger remnant depend strongly on
the mass ratio of the progenitors.  Major galaxy mergers with mass
ratios in the range 1:1 -- 3:1 lead to the formation of boxy or disky
elliptical galaxies, mergers with mass ratios in the intermediate
range 4:1 -- 10:1 result in peculiar galaxies with the morphology of
disk galaxies but kinematics closer to that of elliptical systems, and
minor mergers with mass ratios below 10:1 result in disturbed spiral
galaxies. 

The properties of the merger remnant are also determined by
the gas fraction of the progenitors.  When the gas fraction of the
progenitors is low then the remnants structurally and kinematically
resemble elliptical galaxies.  If the progenitor galaxies are gas-rich
then a prominent pre-existing disc can survive, that is, both major and
minor mergers can produce a disc-dominated galaxy.
\citet{RodriguezGomez2017} considered the influence of mergers on the
galaxy morphology using the Illustris simulation. They found that
mergers play a dominant role in shaping the morphology of massive
galaxies and the mergers do not seem to play any significant role in
determining the morphology of galaxies with masses below
$\sim$10$^{11}M_{\odot}$.

To summarize, the result (remnant) of an interaction or merger is
dependent on a set of parameters such as the mass ratio of
progenitors, the gas fraction in the progenitors, the geometry of the
collision, etc.  One may expect that some specific type of
interaction or merger (with a range of properties of the progenitors
and/or with a range of parameters of the collision) results in a large
offset between the position angles of the photometric and kinematic
major axes.  In our current study, we do not find any morphological
signature of interactions or mergers common to all our galaxies.  It
should be noted that the possible disparity between the time scales on
which an offset between the position angles of the photometric and
kinematic major axes and any other sign of an interaction or merger can
occur makes it difficult to establish the combination of the required
effects of an interaction that result in a large offset between the
position angles of the photometric and kinematic major axes.

\subsection{Shape of the isovelocity curves}
%=====================

We do not find a unique shape for the isovelocity curves in our
galaxies with a large $\Delta$PA. Instead, the isovelocity curves of
the measured velocity fields in some galaxies are more or less close
to a set of parabolic curves (the hourglass-like picture for a
rotating disc) while in other galaxies the curves of isovelocities
resemble more closely straight lines.  

There is no correpondence between the shape of the isovelocity
curves and the presence (lack) of the signs of galaxy interactions.
The hourglass-like picture of the isovelocity curves can be found in
both galaxies with (for example, M-8483-06103) and without (for example,
M-8146-09102) interaction signatures.  Straight-line isovelocity
curves can be also found in galaxies with (for example, M-8547-09102) and
without (for example, M-8140-09101) signs of interactions.  

\subsection{Stellar versus gas rotation}
%=====================

%====================================    Fig  No 4    PAgas vs PAstar 
\begin{figure}
\centering
\resizebox{1.00\hsize}{!}{\includegraphics[angle=000]{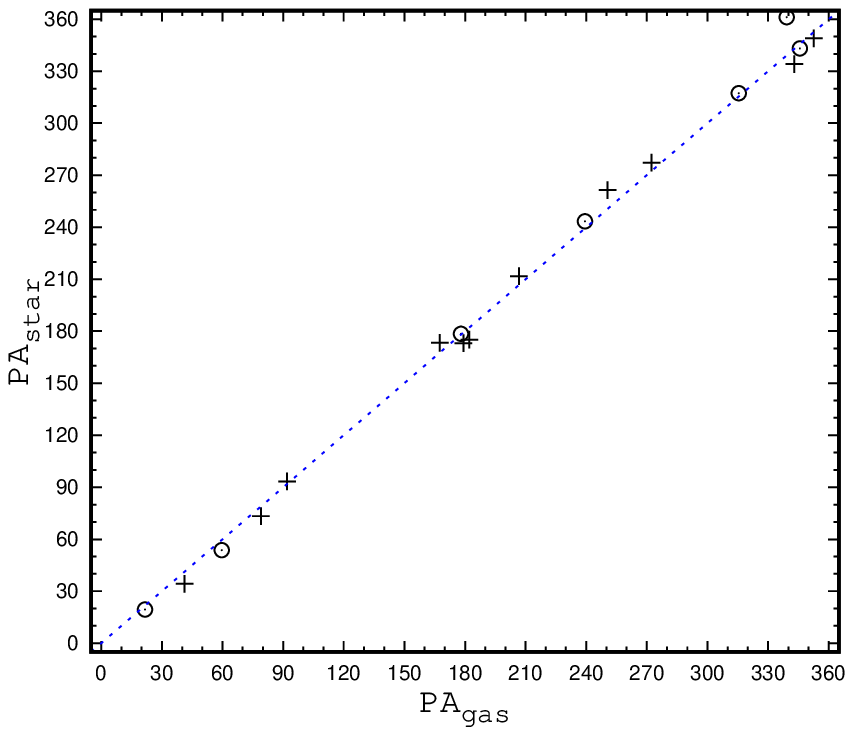}}
\caption{
Position angle of the major axis of the stellar rotation PA$_{star}$
versus position angle of the major axis of the gas rotation PA$_{gas}$
for our sample of the MaNGA galaxies.  The circles stand for data for
galaxies without signs of an interaction, and the plus signs denote
galaxies with signs of an interaction.  The line represents one-to-one
correspondence between these tracers of rotation.
}
\label{figure:pagas-pastar}
\end{figure}

Fig.~\ref{figure:pagas-pastar} shows the position angle of the major
axis of the stellar rotation, PA$_{star}$, as a function the position
angle of the major axis of the gas rotation, PA$_{gas}$.  Inspection
of Fig.~\ref{figure:pagas-pastar} and Table \ref{table:sample} 
shows that the difference between the position angles of the gas and
stellar rotations is within around 10$\degr$ for our galaxies, with
one exception (the difference is 22$\degr$ in the case of the galaxy
M-8140-09101).   

In general, the agreement between the position angles of the gas and
stellar rotations of galaxies is not surprising.
\citet{Chen2016,Jin2016} found that only 10 out of 489 blue MaNGA
galaxies show misaligned gas and stellar rotation by more than
30$\degr$.  \citet{Duckworth2019} found that gas and stellar rotation
to be misaligned by more than 30$\degr$ in only 9 out of 1005
late-type MaNGA galaxies and in 85 out of 204 early-type MaNGA
galaxies.  \citet{Bryant2019} measured the PAs of the stellar and gas
rotation axes in a sample of galaxies from the Sydney-AAO Multi-object
Integral field spectrograph (SAMI) Galaxy Survey.  They found that the
gas and stellar rotation are misaligned by more than 30$\degr$ in
5$\pm$1 per cent in late-type galaxies but the misalignment fraction
is 45$\pm$6 per cent in early-type galaxies.

Thus, our galaxies with large offset between the position angles of
the major kinematic and photometric axes do not show a large offset
between the position angles of the major kinematic gas and stellar
axes, that is, from this point of view, their behavior is similar to the
usual galaxies.

\subsection{Evolution status}
%=====================

%====================================    Fig  No 5     SFR - Msp
\begin{figure}
\centering
\resizebox{1.00\hsize}{!}{\includegraphics[angle=000]{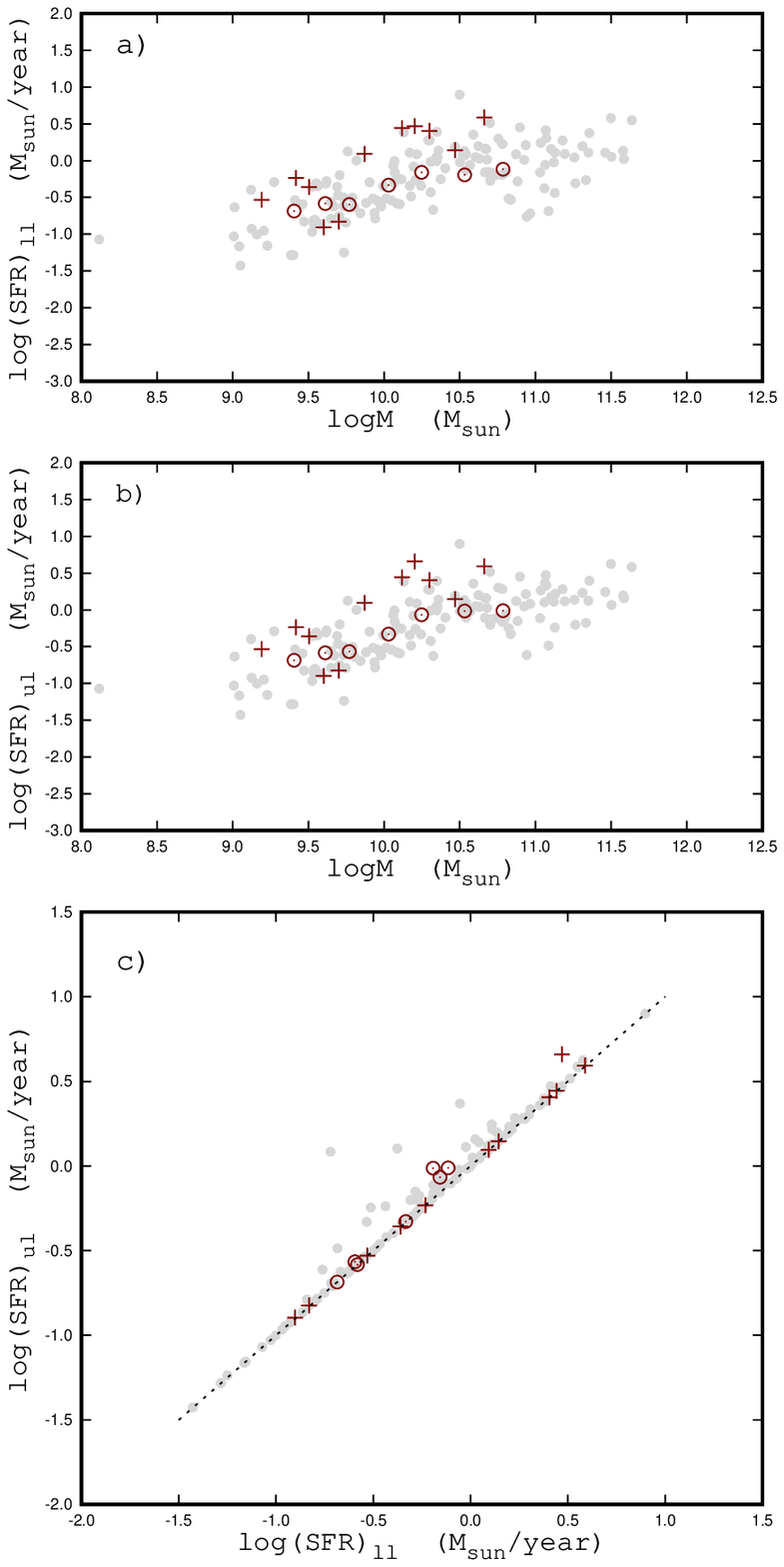}}
\caption{
  {\em Panel} $a$ shows the lower limit of the star formation rate, SFR$_{ll}$, 
  as a function of stellar mass for MaNGA galaxies. The plus signs denote 
  galaxies with $\Delta$PA $> 20\degr$ and with signs of an interaction. 
  The circles denote 
  galaxies with $\Delta$PA $> 20\degr$ and without signs of an interaction. 
  The gray points mark
  galaxies with $\Delta$PA $< 20\degr$ from \citet{Pilyugin2019}.
  {\em Panel} $b$ shows the same as {\em Panel} $a$  but for the upper limit of
  the star formation rate, SFR$_{ul}$.
  {\em Panel} $c$ shows the comparison between the SFR$_{ul}$ and SFR$_{ll}$
  for those galaxies. The line is that of equal values. 
}
\label{figure:m-sfr}
\end{figure}

Many works have been devoted to the study of the star formation rate
vs. stellar mass of galaxies (SFR - $M$) diagram during the last
decade \citep[][among many
others]{Noeske2007,Brinchmann2004,Whitaker2014,Renzini2015,
Lin2017,Belfiore2018,Sanchez2018}.  It was found that majority of
galaxies fall in two distinct bands in this diagram: the band of
galaxies with high SFR, which was named the star-forming main sequence
or simply main sequence, and the band of galaxies with low (if any)
SFR, which is called red and dead, quiescent, or quenched sequence.
The region between these two bands is populated by fewer galaxies,
this region is usually called the green valley.  The position of a
galaxy in the SFR -- $M$ diagram indicates its evolutionary status. 

The star-forming galaxies are located within a rather narrow band in
the SFR -- $M$ diagram.  The dispersion in the SFRs is $\sim$0.2--0.3
dex \citep[e.g.,][]{Speagle2014,Sanchez2018}.  However, the
differences in the mean SFR in star-forming galaxies of a given
stellar mass obtained in different works can be as large as a factor
of two to three depending on the adopted stellar mass and SFR
diagnostics, that is, the assumed stellar initial mass function, the
conversion relation for estimates of the SFR, and the correction for
extinction \citep{Speagle2014}.

Is there any difference in the positions of galaxies with large
$\Delta$PA in the SFR -- $M$ diagram as compared to other star-forming
galaxies?  We estimate the current star formation rate from the
H$\alpha$ luminosity of a galaxy $L_{{\rm H}{\alpha}}$ using the
calibration relation of \citet{Kennicutt1998} reduced by
\citet{Brinchmann2004} for the Kroupa initial mass function
\citep{Kroupa2001}  
\begin{equation}
\log {\rm SFR}  = \log L_{{\rm H}{\alpha}} -41.28 . 
\label{equation:sfr}
\end{equation}
Firstly, the  H$\alpha$ luminosity of a galaxy $L_{{\rm H}{\alpha}}$
was determined as the sum of the H$\alpha$ luminosities of the spaxels
with  H\,{\sc ii}-region-like spectra only. The spaxels with AGN-like
and intermediate spectra (located in the BPT diagram right (above) the
separation curve of \citet{Kauffmann2003}) were rejected.  Since the
H$\alpha$ flux from spaxels with AGN-like and intermediate spectra may
contain a star-forming component \citep[e.g.,][]{Belfiore2018}, the
SFR estimated for the H$\alpha$ luminosities of spaxels with H\,{\sc
ii}-region-like spectra only can be considered as a lower limit of the
star formation rate, SFR$_{ll}$. Further, the MaNGA spectroscopic
measurements do not extend to optical radius for many galaxies. In
those cases, our determinations of the star formation rates are
affected by this finite extent of the available data. However, the
aperture corrections for MaNGA galaxies are small and not applying
them has no impact on the determined star formation rates
\citep{Belfiore2018}.

Panel $a$ of Fig.~\ref{figure:m-sfr} shows the SFR$_{ll}$ as a
function of stellar mass for MaNGA galaxies.  The plus signs
denote galaxies with $\Delta$PA $> 20\degr$ and with the signs of
interactions considered here.  The circles denote galaxies with
$\Delta$PA $> 20\degr$ and without interaction signatures. The gray
points mark galaxies with $\Delta$PA $< 20\degr$ from
\citet{Pilyugin2019}.

Next, the H$\alpha$ luminosity of our galaxies, $L_{{\rm H}{\alpha}}$,
is determined as the sum of the H$\alpha$ luminosities of all the
spaxels.  The SFR estimated in such a way can be considered as an
upper limit of the present-day star formation rate, SFR$_{ul}$.  Panel
$b$ of Fig.~\ref{figure:m-sfr} shows the SFR$_{ul}$ as a function of
stellar mass for our MaNGA galaxies.  Panel $c$ of
Fig.~\ref{figure:m-sfr} shows the SFR$_{ul}$ as a function of
SFR$_{ll}$.   Again, the plus signs denote galaxies with
$\Delta$PA $> 20\degr$ and with signs of an interaction, the circles
show galaxies with $\Delta$PA $> 20\degr$ and without visible
interaction signatures, and the gray points mark galaxies with
$\Delta$PA $< 20\degr$.  

Inspection of Fig.~\ref{figure:m-sfr} shows that the spaxels with
H\,{\sc ii}-region-like spectra provide a dominant contribution to the
H$\alpha$ luminosity in the bulk of the galaxies of our sample.  The
spaxels with AGN-like and intermediate spectra make an appreciable
contribution to the H$\alpha$ luminosity in massive galaxies that are
AGN hosts.  The positions of those galaxies in the SFR -- $M$ diagram
are shifted towards the green valley. This is in agreement with
results from \citet{Sanchez2018}.

Examination of Fig.~\ref{figure:m-sfr} shows that the galaxies with
$\Delta$PA $> 20\degr$ are located in the same area in the SFR -- $M$
diagram as the galaxies with $\Delta$PA $< 20\degr$, that is, the
galaxies with $\Delta$PA $> 20\degr$ lie within the main sequence
outlined by galaxies with $\Delta$PA $< 20\degr$.  However,
galaxies with signs of an interaction show on average higher SFRs in
comparison to galaxies of the same mass without interaction
signatures.

\subsection{Central oxygen abundance}
%=====================

%====================================    Fig  No 6   OH gradients in galaxies
\begin{figure*}
\resizebox{1.00\hsize}{!}{\includegraphics[angle=000]{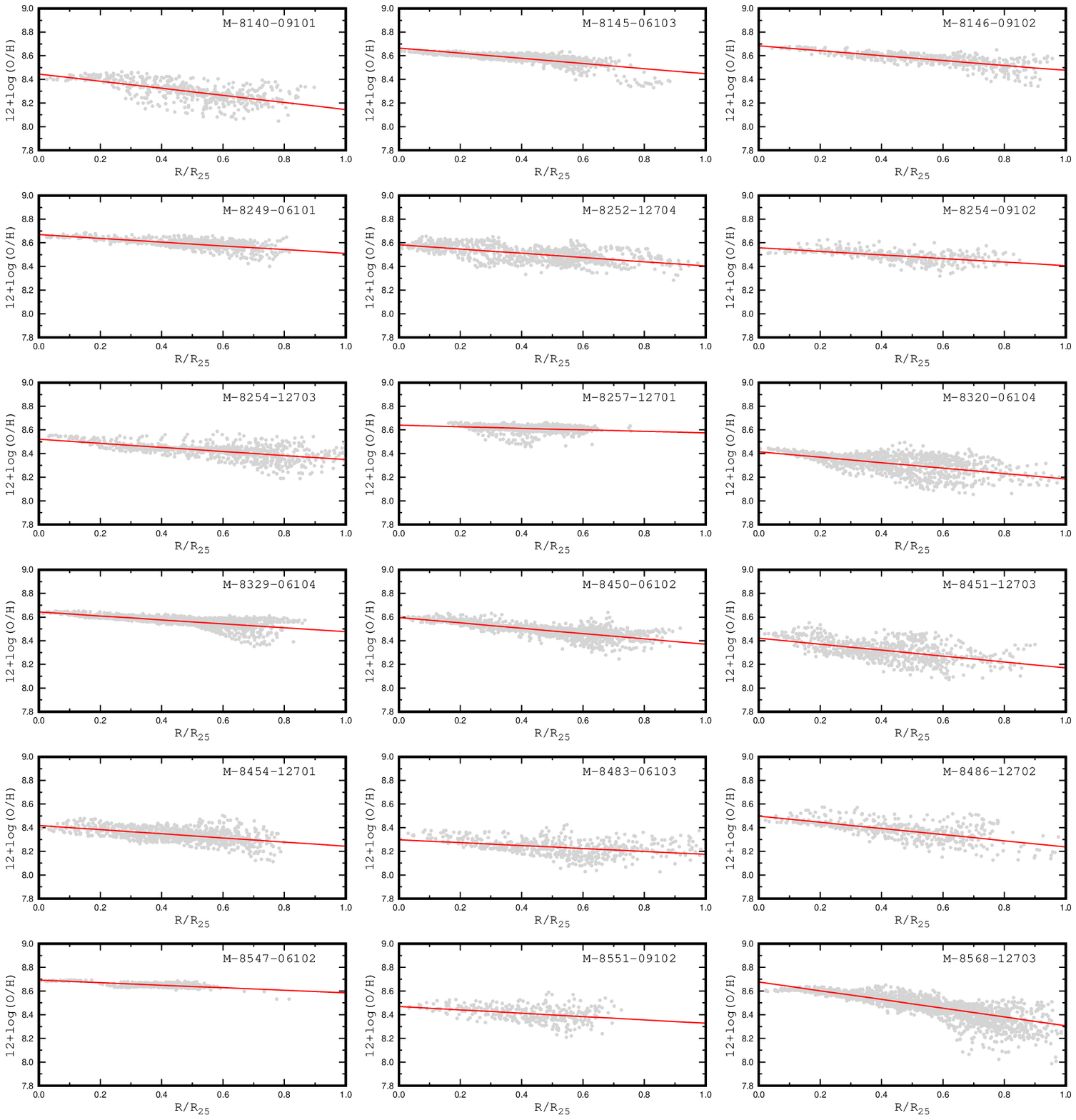}}
\caption{
Radial distributions of the oxygen abundances for our sample of MaNGA
galaxies.  The grey points indicate the oxygen abundances in
individual spaxels.  The solid line represents the inferred linear
abundance gradient.
}
\label{figure:rg-oh}
\end{figure*}

%====================================    Fig  No 7        Msp - OHo diagram  
\begin{figure}
\resizebox{1.00\hsize}{!}{\includegraphics[angle=000]{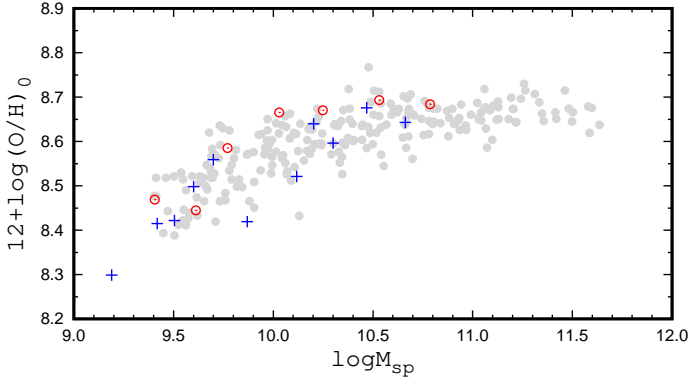}}
\caption{
Central intersect oxygen abundance as a function of spectroscopic
mass of a galaxy.  The galaxies with a large offest between the
position angles of the major kinematic and photometric axes
($\Delta$PA $> 20\degr$) are shown by plus signs (for galaxies with
signs of interactionis) and by circles (for seemingly non-interacting
galaxies).  The grey points denote MaNGA galaxies with $\Delta$PA $<
20\degr$ from \citet{Pilyugin2019}.
}
\label{figure:m-oho}
\end{figure}

%====================================    Fig  No 8        dOHo vs dPA and Msp
\begin{figure}
\centering
\resizebox{1.00\hsize}{!}{\includegraphics[angle=000]{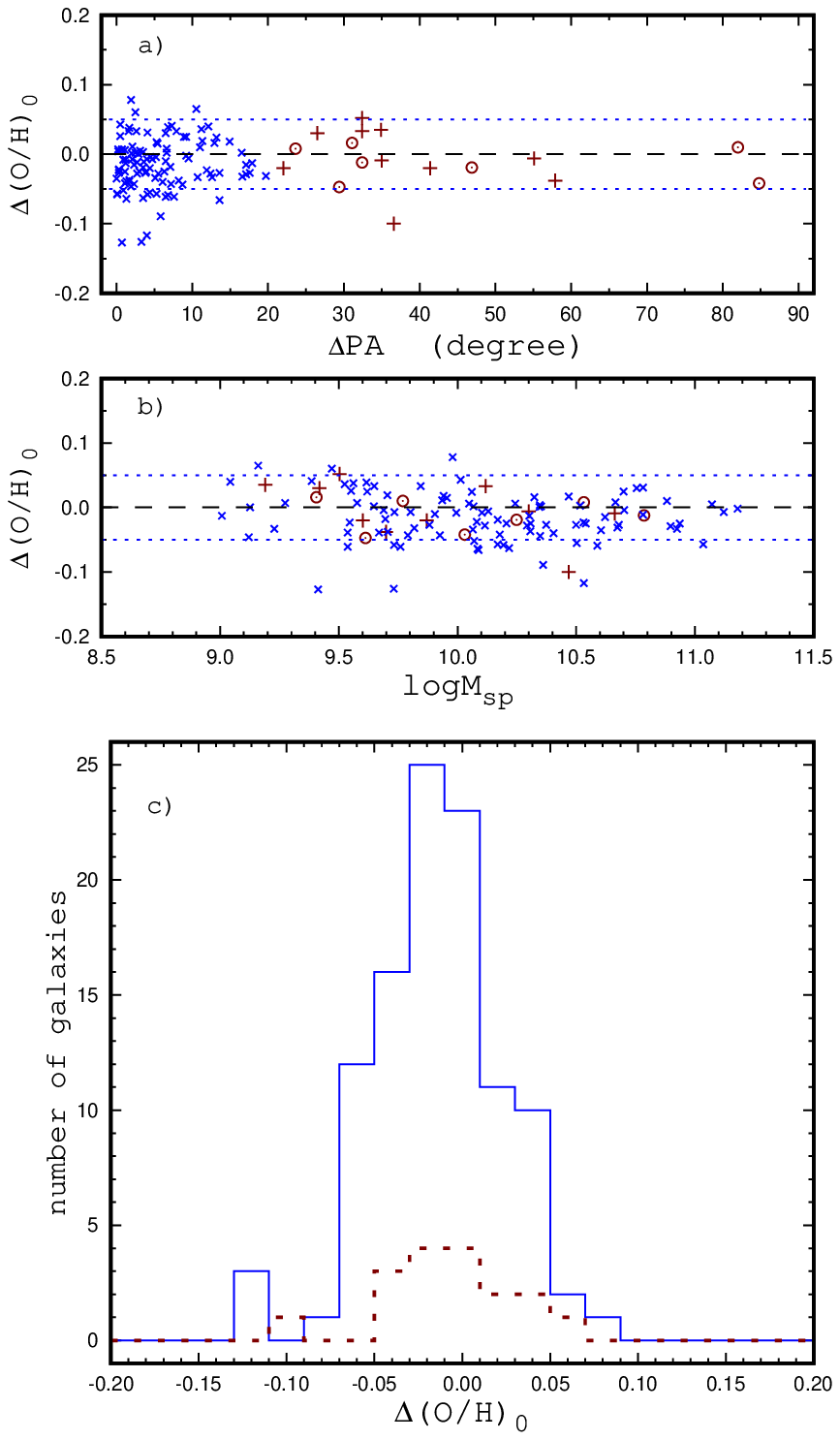}}
\caption{
Value of depletion (enhancement) of the central oxygen abundance
$\Delta$(O/H)$_{0}$ (the difference between the mean value of the
oxygen abundances 12+log(O/H) in spaxels with galactocentric distances
within $0.1R_{25}$ and the value of the central intersect oxygen
abundance estimated from the radial abundance gradient based on
spaxels with galactocentric distances within 0.2 -- $0.8R_{25}$) as a
function of $\Delta$PA  ({\em panel} $a$) and as a function of stellar
mass of the galaxy ({\em panel} $b$). The circles represent data for
galaxies with  large $\Delta$PA and without signs of ah interaction
while the plus signs show galaxies with large $\Delta$PA and with
signs of an interaction.  The crosses denote galaxies with small (no)
$\Delta$PA from \citet{Pilyugin2019}.  The dashed line marks a
$\Delta$PA of zero, and the dotted lines show the $\pm$0.05 values.
{\em Panel} $c$ shows histograms of $\Delta$(O/H)$_{0}$ for galaxies
with $\Delta$PA $< 20\degr$ (solid line) and for galaxies with
$\Delta$PA $> 20\degr$ (dashed line). 
}
\label{figure:doho}
\end{figure}

The radial oxygen abundance distribution in a galaxy is traditionally 
described by a straight line of the form 
\begin{equation}
{\rm (O/H)}^{*} = {\rm (O/H)}_{0}^{*} + grad \, \times \, R_{g}
\label{equation:gradient}
\end{equation}
where (O/H)$^{*}$ $\equiv$ 12 + log(O/H)($R$) is the oxygen abundance
at the fractional radius $R_{g}$ (normalized to the optical radius
$R_{25}$), (O/H)$_{0}^{*}$ $\equiv$ 12 + log(O/H)$_{0}$ is the
intersect central oxygen abundance, and $grad$ is the slope of the
oxygen abundance gradient expressed in terms of dex/$R_{25}$.
Fig.~\ref{figure:rg-oh} shows the radial distributions of the oxygen
abundances for our sample of MaNGA galaxies.  The grey points indicate
the oxygen abundances in individual spaxels.  The solid line
represents the inferred linear abundance gradient.  The galactocentric
distance of each spaxel was estimated using the position of the
photometric major axis and the photometric inclination angle.
Inspection of Fig.~\ref{figure:rg-oh} shows that the radial abundance
distribution in the majority of galaxies can be reproduced rather well
by a straight line. 

Fig.~\ref{figure:m-oho} shows the comparison of the (O/H)$_{0}$ -- $M$
diagrams for galaxies with large ($\Delta$PA $\ga 20\degr$) and small
(no) offests between the position angles of the major kinematic and
photometric axes.  The plus signs stand for galaxies with large
$\Delta$PA and with signs of an interaction.  The circles denote the
galaxies with large $\Delta$PA and without signs of an interaction.  
The grey points indicate MaNGA galaxies with small (no) offests
between the position angles of the major kinematic and photometric
axes from \citet{Pilyugin2019}.  Inspection of Fig.~\ref{figure:m-oho}
shows that the central intersect oxygen abundances in the galaxies
with large and small $\Delta$PA are located in the same area in the
(O/H)$_{0}$ -- $M$ diagram.

We estimate the depletion in the oxygen abundance at the center of a
galaxy in the following way. The radial oxygen abundance gradient is
determined based on the spaxels with galactocentric distances from
0.2$R_{g}$ to 0.8$R_{g}$.  This excludes the influence of possible
depletions (enhancements) in the oxygen abundance at the center and at
the periphery of the galaxy on the obtained radial abundance gradient
and on the central intersect value of the oxygen abundance
(O/H)$_{0}$. The local central oxygen abundance (O/H)$_{C}$ is
estimated as the average value of the oxygen abundances in spaxels
with galactocentric distances within 0.1$R_{g}$.  The depletion
(enhancement) of the central oxygen abundance $\Delta$(O/H)$_{0}$ is
defined as difference between the local and intersect central oxygen
abundances $\Delta$(O/H)$_{0}$ = log(O/H)$_{C}$ -- log(O/H)$_{0}$. 

Panel $a$ of Fig.~\ref{figure:doho} shows the value of the depletion
(enhancement) of the central oxygen abundance $\Delta$(O/H)$_{0}$ as a
function of the absolute value of the offset between the kinematic and
photometric major axes $\Delta$PA  = $|$PA$_{kin}$ - PA$_{phot}$$|$.
The circles mark galaxies with $\Delta$PA $\ga$ 20$\degr$ and
without signs of an interaction.  The plus signs denote galaxies with
$\Delta$PA $\ga$ 20$\degr$ and with interaction signatures.  The
crosses are MaNGA galaxies with small offsets between the kinematic
and photometric position angles ($\Delta$PA $\la$ 20$\degr$) from
\citet{Pilyugin2019}.   Panel $b$ of Fig.~\ref{figure:doho} shows the
$\Delta$(O/H)$_{0}$ as a function of the stellar mass.  Panel $c$ of
Fig.~\ref{figure:doho} shows histograms of $\Delta$(O/H)$_{0}$ for
galaxies with $\Delta$PA $< 20\degr$ (solid line) and for galaxies
with $\Delta$PA $> 20\degr$ (dashed line).  Inspection of
Fig.~\ref{figure:doho} shows that the depletion (enhancement) of the
central oxygen abundance is within 0.05 dex for the majority of
galaxies with both large and small offests between the kinematic and
photometric position angles. 

Examination of panel $c$ of Fig.~\ref{figure:doho} shows that there is
a systematic depletion by $\sim$0.02 dex in the local central oxygen
abundances in our sample of MaNGA galaxies.  This small systematic
depletion in the local central oxygen abundances can be false.
\citet{Belfiore2017} have examined the influence of the point spread
function (PSF) of the MaNGA measurements on oxygen abundance
determinations. They found that the influence of the PSF on the
obtained value of the oxygen abundance is maximum at the center of the
galaxy in the sense that the obtained oxygen abundance may be
underestimated by about 0.04 dex or less depending on the value of the
inclination angle and the ratio between the effective raius of the
galaxy and the full width at the half maximum of the PSF.  The value
of the systematic depletion of $\sim$0.02 dex in the local central
oxygen abundances in our sample of MaNGA galaxies is well within the
predictions by\citet{Belfiore2017}.  If this depletion is real then
this suggests that there is a low-rate gas infall into the centers of
those spiral galaxies. 

\subsection{Radial abundance gradient}
%=====================

%====================================    Fig  No 9     OH gradient -- Msp 
\begin{figure}
\resizebox{1.00\hsize}{!}{\includegraphics[angle=000]{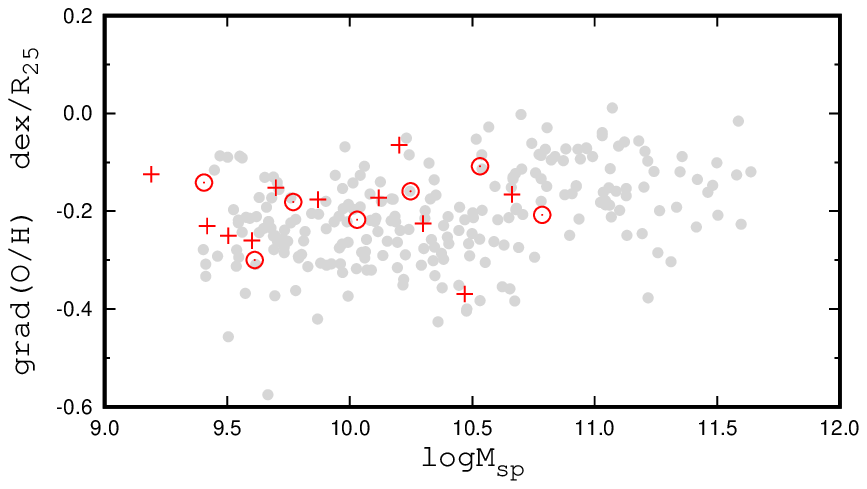}}
\caption{
Radial oxygen abundance gradient in units of dex/$R_{25}$ in the
discs of our galaxies as a function of their spectroscopic mass.  The
galaxies with $\Delta$PA $> 20\degr$ are shown by circles (galaxies
without signs of an interaction) and by plus signs (galaxies with
signs of an interaction).  The grey points are galaxies with
$\Delta$PA $< 20\degr$ from \citet{Pilyugin2019}.
}
\label{figure:m-grad}
\end{figure}

Numerical simulations predict that gas flows induced by interactions
redistribute gas in such a manner that the radial abundance gradients
in the galactic disks flatten during the interacting or merger
\citep[e.g.,][]{Rupke2010a,Rupke2010b,Rich2012,Bustamante2018}.

Fig.~\ref{figure:rg-oh} shows the radial distributions of the oxygen
abundances for our sample of MaNGA galaxies.  Fig.~\ref{figure:m-grad}
shows the radial abundance gradient in the galaxies as a function of
their spectroscopic mass.  The circles stand for galaxies with
$\Delta$PA $\ga 20\degr$ and without signs of an interaction.  The
plus signs denote galaxies with $\Delta$PA $\ga 20\degr$ and with
interaction signatures.   The grey points are individual MaNGA
galaxies with small (no) offests between the kinematic and photometric
position angles ($\Delta$PA $\la 20\degr$) from \citet{Pilyugin2019}. 

Examination of Fig.~\ref{figure:m-grad} shows that the galaxies with
large $\Delta$PA lie within the area outlined by the galaxies with
small $\Delta$PA.

\subsection{Discussion}
%=====================

The position angle of the major photometric axis and other photometric
characteristics of a galaxy are defined by the spatial orientation of
the galaxy and by the distribution of the positions of the stars and
gas within the galaxy.  Similarly, the position angle of the major
kinematic axis and other kinematic characteristics of a galaxy are
defined by the spatial orientation of the galaxy and by the
distribution of the velocities of the stars (and gas) within the
galaxy.  Since the stellar and gas velocity fields in our sample of
galaxies are close to each other (see above) and since the precision
of the measurements of gas velocities is higher than that for stars,
we consider the gas velocies below.  One can say that the large offset
between the position angles of the photometric and kinematic major
axes demonstrates that the measured distributions of positions and
velocities of the stellar component are in conflict to each other,
that is, this unusual property is common to all those galaxies. 

However, other properties of our sample of galaxies with large
$\Delta$PA are rather usual.  There is no other characteristic that is
common for all those galaxies, that is, there is no parameter that has a
similar value for all those galaxies and can serve as an additional
indicator of galaxies with large $\Delta$PA.  This suggests that
either the reason responsible for the large  $\Delta$PA does not
influence other characteristics of a galaxy or that the galaxies with
large  $\Delta$P are not uniform in their origin; instead they may
have different evolutionary pathways. 

It is interesting to note that there is a possibility that large
$\Delta$PA can be observed in a galaxy even if the real distribution
of positions and velocities of stars within that galaxy are in
agreement with each other but the observed line-of-sight velocity
field is distorted.  One may assume that the observed line-of-sight
velocity field of a galaxy with a large $\Delta$PA is a composition of
two velocity fields (two components of the galaxy motion).  The first
component of the galaxy motion is the usual disk rotation where the
major kinematic axis coincides with the major photometric axis.  The
second component of the galaxy motion is a rotation of
the galaxy as a whole around some axis that does not coincide with the
disc rotation axis.  This concept can explain the observed offsets
between the measured position angles of the photometric and kinematic
major axes in galaxies.

\section{Conclusions}
%=====================

We derive the photometric, kinematic, and abundance properties of 18
star-forming MaNGA galaxies with fairly regular velocity fields and
surface brightness distributions and with a large offset between the
measured position angles of the major kinematic and photometric axes,
$\Delta$PA $\ga 20\degr$. We aim to examine if there is any other
distinctive characteristic common to all these galaxies.

We found the following properties for those galaxies. \\ 
-- Morphological signs of interactions or mergers can be found in some
(in 11 out of 18) but not in all galaxies with large $\Delta$PA. \\
 -- There is no unique shape for the isovelocity curves in the
galaxies with large $\Delta$PA. The isovelocity curves in the measured
velocity fields in some galaxies are close to parabolic curves (that is,
the hourglass-like picture for a rotating disk), while in other
galaxies the isovelocity curves closely resemble straight lines.  There
is no correpondence between the shape of the isovelocity curves and
the presence (lack) of interaction signatures in a galaxy.  The
hourglass-like picture of the isovelocity curves can be found both in
galaxies with and without signs of an interaction.  The isovelocities
of the straight line shape can be also found both in galaxies with and
without signs of an interaction.   \\
--  The position angle of the gas rotation coincides with (or at least
is close to) the position angle of the stellar rotation. \\ 
-- The positions of the galaxies with a large $\Delta$PA in the star
formation rate versus stellar mass diagram show that those galaxies
belong to the main sequence of the star-forming galaxies.  
However, galaxies with signs of an interaction show an on average
higher SFR in comparison to the galaxies of the same mass without
signs of an interaction.    \\ 
-- The oxygen abundances (and, consequently, astration levels) are
rather similar for our samples of galaxies with large and small (no)
$\Delta$PA in the sense that the locations of the galaxies with a
large $\Delta$PA in the central oxygen abundance versus stellar mass
diagram are within (and distributed across the whole) area outlined by
the galaxies with small $\Delta$PA.  The positions of the galaxies
with a large $\Delta$PA in the radial abundance gradient versus
stellar mass diagram are also within the area outlined by the galaxies
with small $\Delta$PA. 

Thus, there is not any distinctive characteristic common to all the
galaxies with large $\Delta$PA; the considered properties of the
galaxies with a large $\Delta$PA are rather similar to those of
galaxies with small (no) $\Delta$PA.  This suggests that either the
reason responsible for the large  $\Delta$PA does not influence those
properties of a galaxy or the galaxies with a large  $\Delta$PA are
not uniform in their origin, they can, instead, originate from different
evolutionary pathways.

%==========================
\section*{Acknowledgements}

We are grateful to the referee for his/her constructive comments. \\
L.S.P., E.K.G., and I.A.Z.\  acknowledge support within the framework
of Sonderforschungsbereich (SFB 881) on ``The Milky Way System''
(especially subproject A5), which is funded by the German Research
Foundation (DFG). \\ 
L.S.P.\ and I.A.Z.\ thank for hospitality of the
Astronomisches Rechen-Institut at Heidelberg University, where part of
this investigation was carried out. \\
L.S.P.\  acknowledges financial support from the State Agency for Research
of the Spanish MCIU through the ``Center of Excellence Severo Ochoa''
award to the Instituto de Astrof\'{i}sica de Andaluc\'{i}a (SEV-2017-0709). \\ 
%I.A.Z.\ acknowledges the support of the Volkswagen Foundation 
%under the Trilateral Partnerships grant No.\ 90411. \\
I.A.Z. acknowledges the support from the National Academy of Sciences of Ukraine 
by the project 417Kt. \\
J.M.V acknowledges financial support from projects AYA2017-79724-C4-4-P,
of the Spanish PNAYA, and Junta de Andalucia Excellence PEX2011-FQM705. \\
This work was partly funded by the subsidy allocated to Kazan Federal 
University for the state assignment in the sphere of scientific 
activities (L.S.P.).  \\ 
We acknowledge the usage of the HyperLeda database (http://leda.univ-lyon1.fr). \\
Funding for SDSS-III has been provided by the Alfred P. Sloan Foundation,
the Participating Institutions, the National Science Foundation,
and the U.S. Department of Energy Office of Science.
The SDSS-III web site is http://www.sdss3.org/. \\
Funding for the Sloan Digital Sky Survey IV has been provided by the
Alfred P. Sloan Foundation, the U.S. Department of Energy Office of Science,
and the Participating Institutions. SDSS-IV acknowledges
support and resources from the Center for High-Performance Computing at
the University of Utah. The SDSS web site is www.sdss.org. \\
SDSS-IV is managed by the Astrophysical Research Consortium for the 
Participating Institutions of the SDSS Collaboration including the 
Brazilian Participation Group, the Carnegie Institution for Science, 
Carnegie Mellon University, the Chilean Participation Group,
the French Participation Group, Harvard-Smithsonian Center for Astrophysics, 
Instituto de Astrof\'isica de Canarias, The Johns Hopkins University, 
Kavli Institute for the Physics and Mathematics of the Universe (IPMU) / 
University of Tokyo, Lawrence Berkeley National Laboratory, 
Leibniz Institut f\"ur Astrophysik Potsdam (AIP),  
Max-Planck-Institut f\"ur Astronomie (MPIA Heidelberg), 
Max-Planck-Institut f\"ur Astrophysik (MPA Garching), 
Max-Planck-Institut f\"ur Extraterrestrische Physik (MPE), 
National Astronomical Observatories of China, New Mexico State University, 
New York University, University of Notre Dame, 
Observat\'ario Nacional / MCTI, The Ohio State University, 
Pennsylvania State University, Shanghai Astronomical Observatory, 
United Kingdom Participation Group,
Universidad Nacional Aut\'onoma de M\'exico, University of Arizona, 
University of Colorado Boulder, University of Oxford, University of Portsmouth, 
University of Utah, University of Virginia, University of Washington, University of Wisconsin, 
Vanderbilt University, and Yale University.

%\end{document}

\appendix
%=====================

\section{Maps of the inferred properties of our galaxies}
%=====================

The figures in this Section show the inferred properties of our MaNGA
galaxies.  In each figure, panel $a$ shows the surface brightness
distribution across the image of the galaxy in sky coordinates
(pixels). North is up and east is left.  The pixel scale is 0.5
arcsec. The value of the surface brightness is color-coded. The
circle shows the photometric center of the galaxy. The solid line
indicates the position of the major photometric axis of the galaxy.
The plus sign denotes the kinematic center of the galaxy determined
from the H${\alpha}$ velocity field. The dashed line indicates the
position of the major kinematic  (H${\alpha}$) axis of the galaxy.
Panel $b$ shows the color-coded observed (line of sight) H${\alpha}$
velocity field of a given galaxy in sky coordinates.    Panel $c$
shows the color-coded observed stellar velocity field.  Panel $d$
shows the BPT diagram. The symbols are individual spaxels.  Solid and
long-dashed curves mark the demarcation line between AGNs and H\,{\sc
ii} regions defined by \citet{Kauffmann2003} and \citet{Kewley2001},
respectively.  The short-dashed line is the dividing line between
Seyfert galaxies and LINERs defined by \citet{CidFernandes2010}.  The
black points are AGN-like objects according to the dividing line of
\citet{Kewley2001}.  The blue points are H\,{\sc ii}-region-like
objects according to the dividing line of \citet{Kauffmann2003}.  The
red points are intermediate objects located between the dividing lines
of \citet{Kauffmann2003} and \citet{Kewley2001}.  Panel $e$ shows the
locations of the AGN-like,  H\,{\sc ii}-region-like, and intermediate
spaxels in the image of the galaxy.  Panel $f$ is a plot of the oxygen
abundance distribution across the image of the galaxy in sky
coordinates (pixels). The value of the oxygen abundance is
color-coded.

\clearpage

%====================================    Fig  No  M-8140-09101 
\begin{figure}
\resizebox{1.00\hsize}{!}{\includegraphics[angle=000]{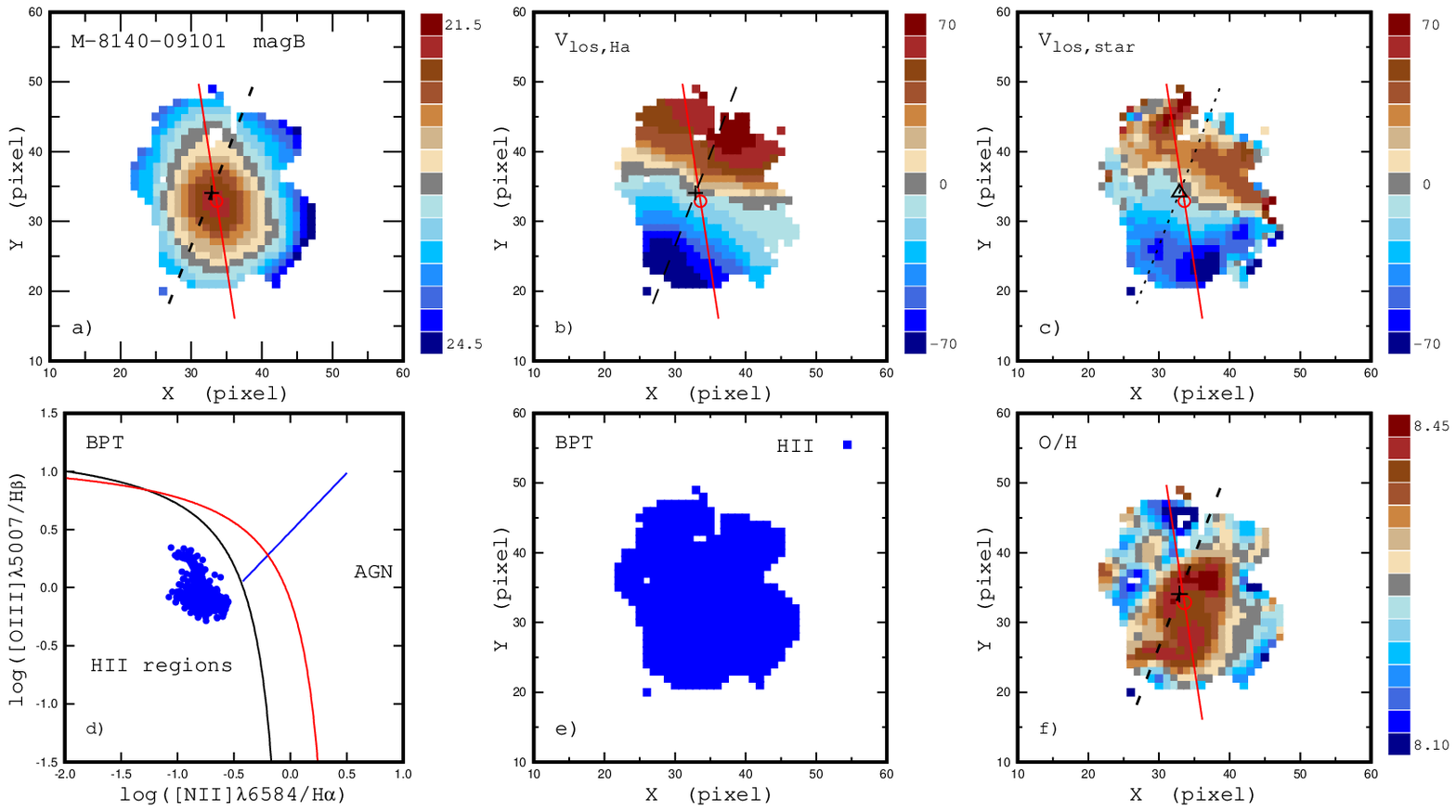}}
\caption{
Inferred properties of the MaNGA galaxy M-8140-09101.
}
\label{figure:m-8140-09101}
\end{figure}

%====================================    Fig  No  M-8145-06103 
\begin{figure}
\resizebox{1.00\hsize}{!}{\includegraphics[angle=000]{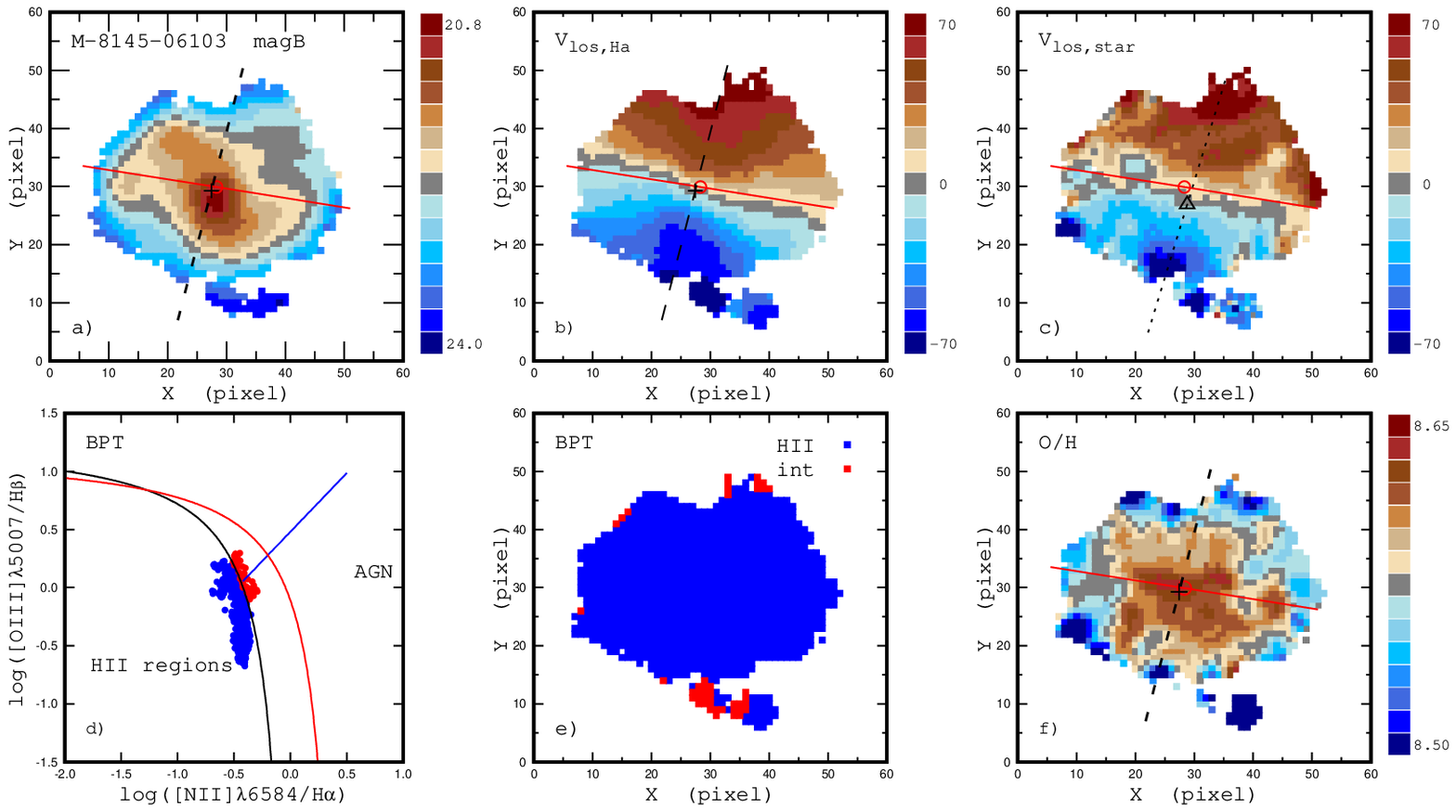}}
\caption{
Inferred properties of the MaNGA galaxy M-8145-06103.
}
\label{figure:m-8145-06103}
\end{figure}

%====================================    Fig  No  M-8146-09102 
\begin{figure}
\resizebox{1.00\hsize}{!}{\includegraphics[angle=000]{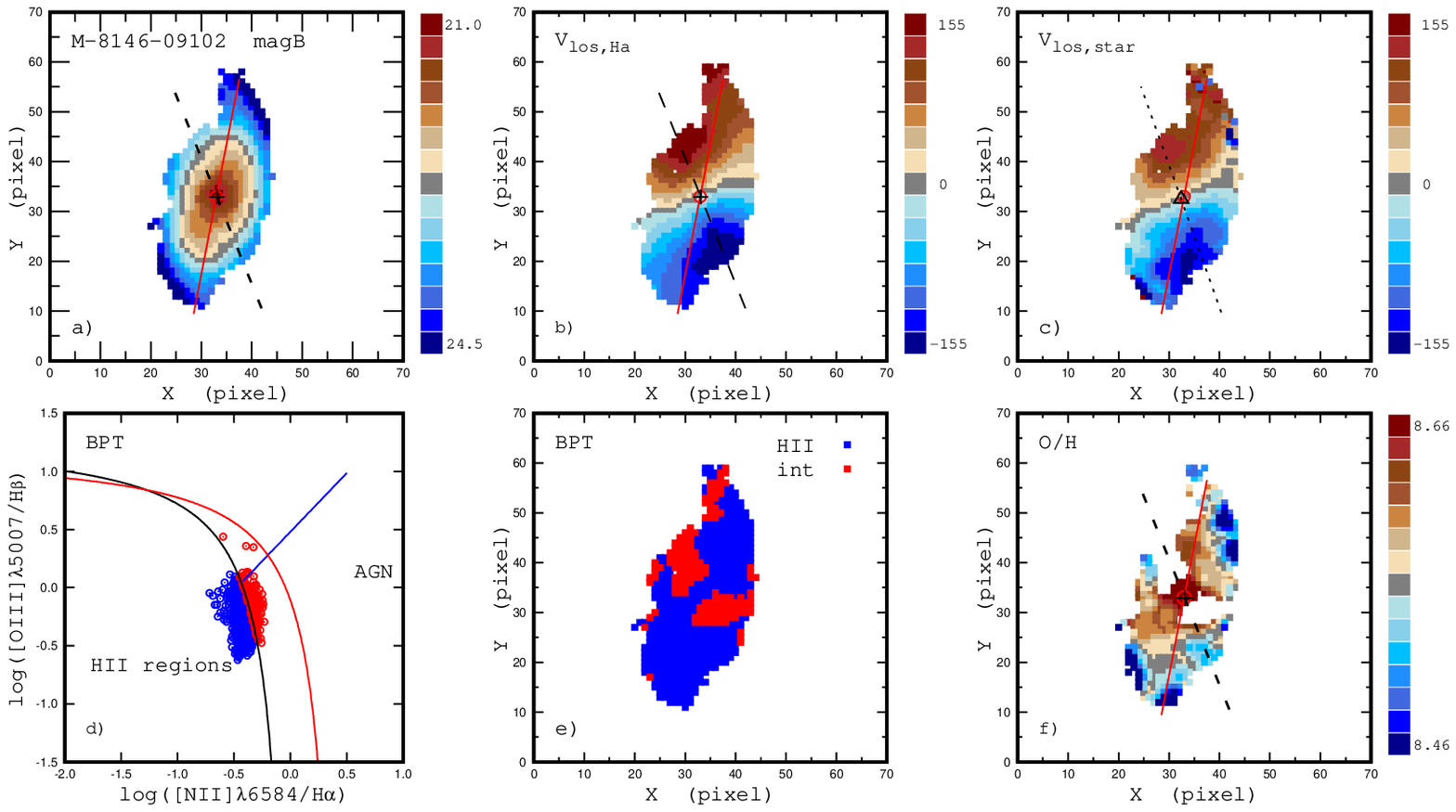}}
\caption{
Inferred properties of the MaNGA galaxy M-8146-09102.
}
\label{figure:m-8146-09102}
\end{figure}

%====================================    Fig  No  M-8249-06101 
\begin{figure}
\resizebox{1.00\hsize}{!}{\includegraphics[angle=000]{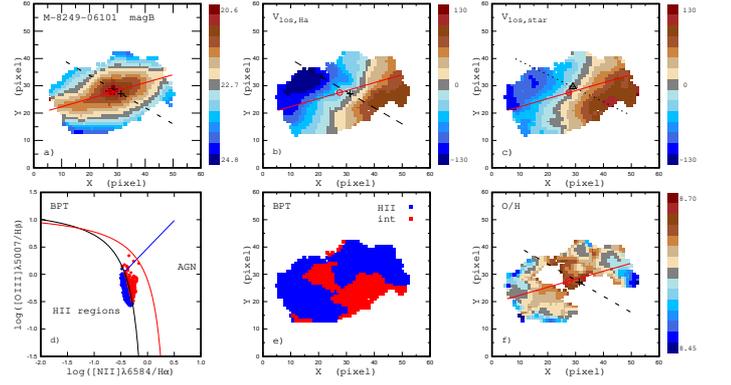}}
\caption{
Inferred properties of the MaNGA galaxy M-8249-06101.
}
\label{figure:m-8249-06101}
\end{figure}

%====================================    Fig  No  M-8252-12704 
\begin{figure}
\resizebox{1.00\hsize}{!}{\includegraphics[angle=000]{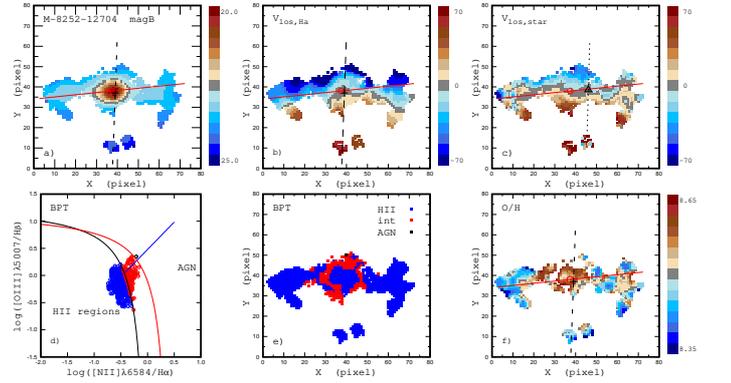}}
\caption{
Inferred properties of the MaNGA galaxy M-8252-12704.
}
\label{figure:m-8252-12704}
\end{figure}

%====================================    Fig  No  M-8254-09102 
\begin{figure}
\resizebox{1.00\hsize}{!}{\includegraphics[angle=000]{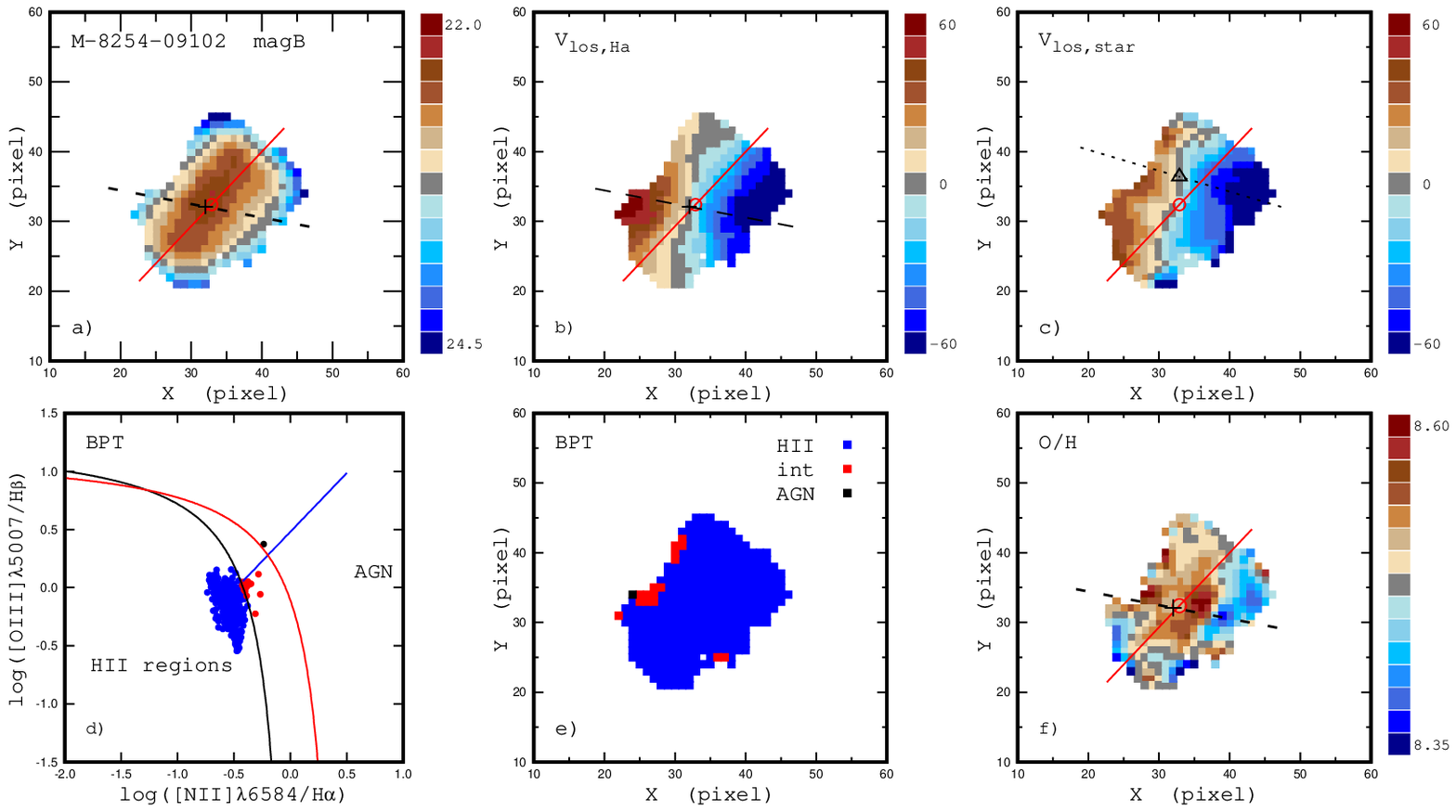}}
\caption{
Inferred properties of the MaNGA galaxy M-8254-09102.
}
\label{figure:m-8254-09102}
\end{figure}

%====================================    Fig  No  M-8254-12703 
\begin{figure}
\resizebox{1.00\hsize}{!}{\includegraphics[angle=000]{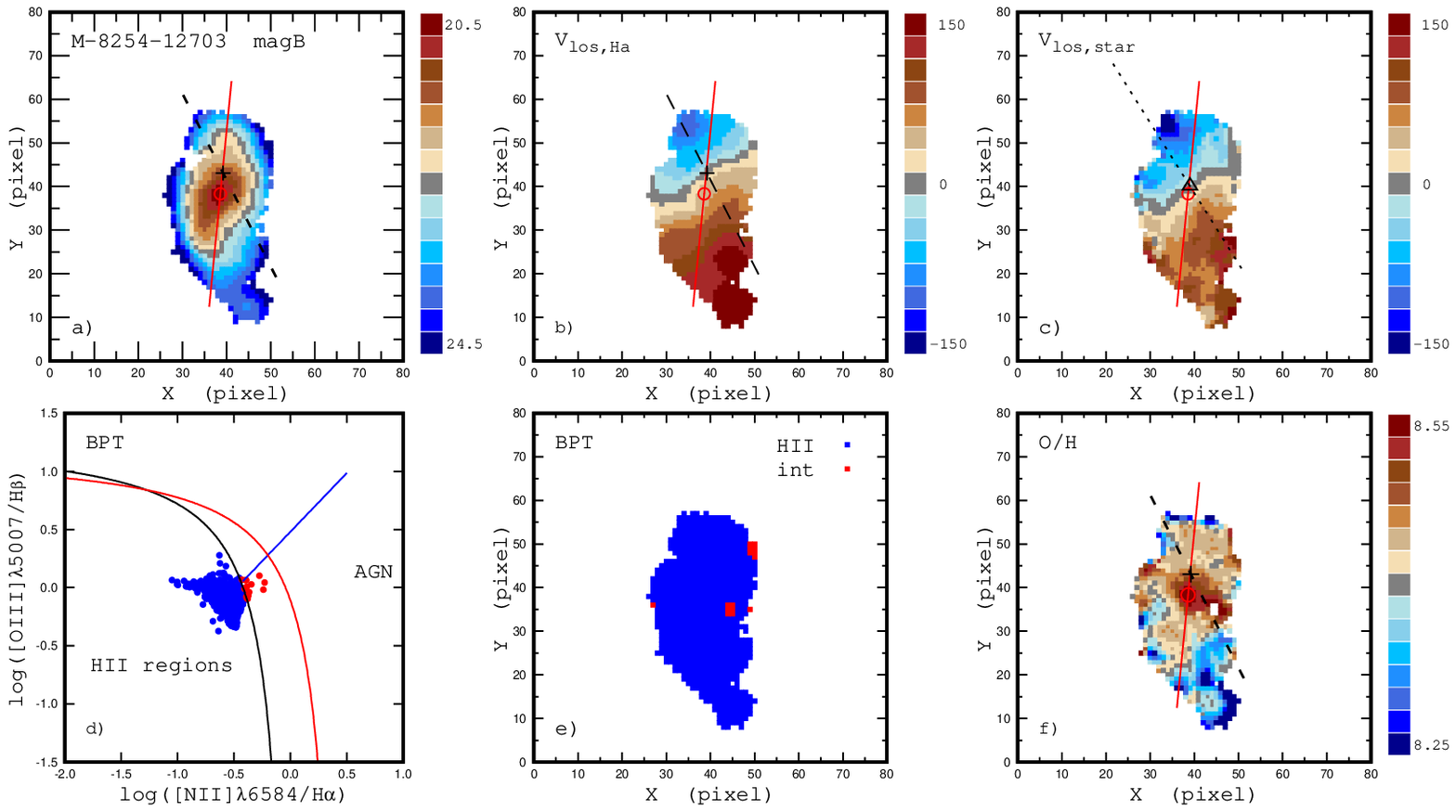}}
\caption{
Inferred properties of the MaNGA galaxy M-8254-12703.
}
\label{figure:m-8254-12703}
\end{figure}

%====================================    Fig  No  M-8257-12701 
\begin{figure}
\resizebox{1.00\hsize}{!}{\includegraphics[angle=000]{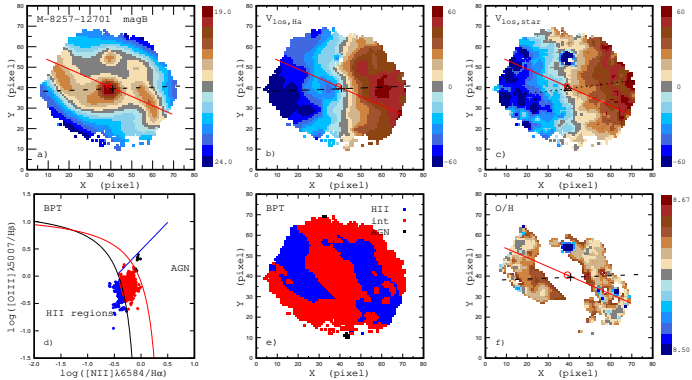}}
\caption{
Inferred properties of the MaNGA galaxy M-8257-12701.
}
\label{figure:m-8257-12701}
\end{figure}

%====================================    Fig  No  M-8320-06104 
\begin{figure}
\resizebox{1.00\hsize}{!}{\includegraphics[angle=000]{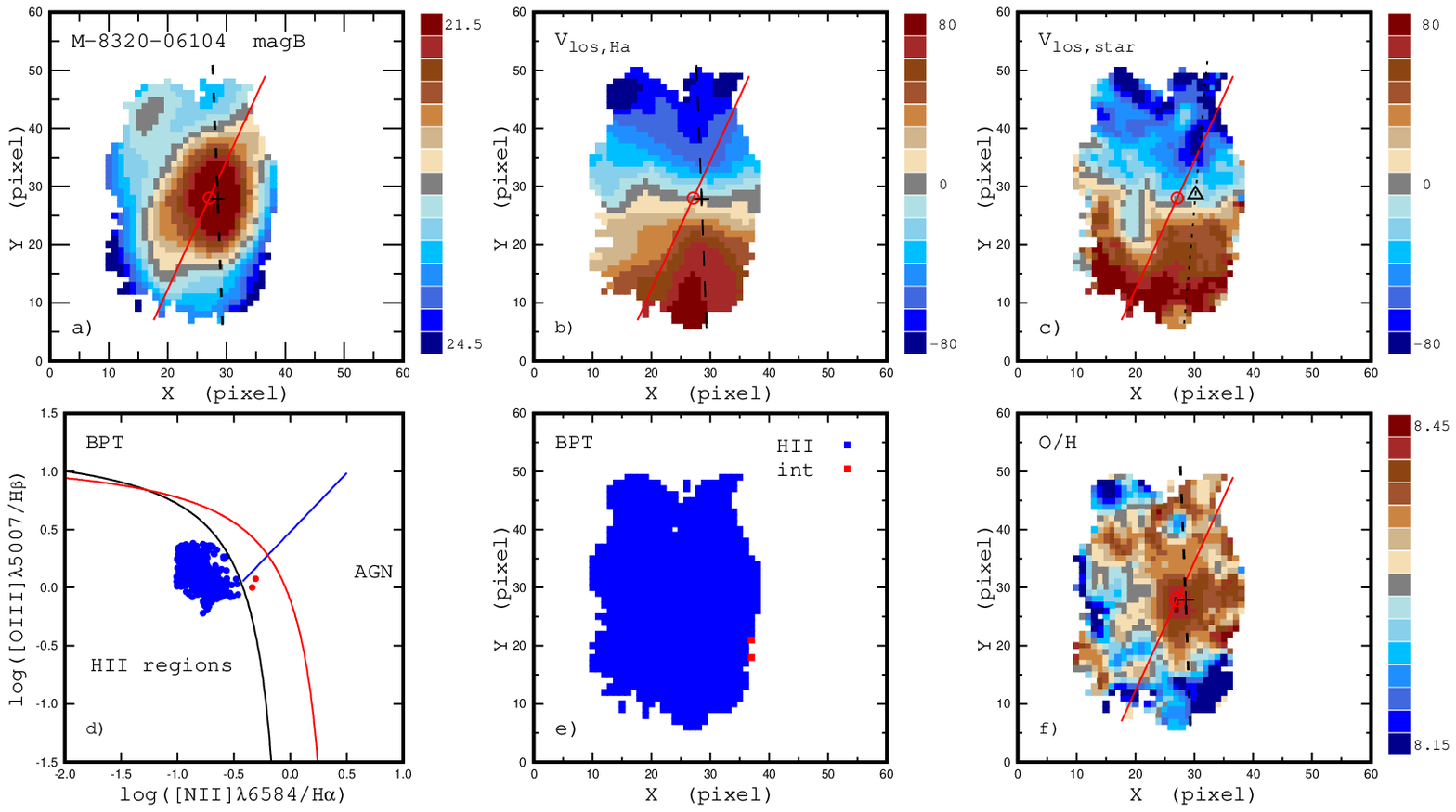}}
\caption{
Inferred properties of the MaNGA galaxy M-8320-06104.
}
\label{figure:m-8320-06104}
\end{figure}

%====================================    Fig  No  M-8329-06104 
\begin{figure}
\resizebox{1.00\hsize}{!}{\includegraphics[angle=000]{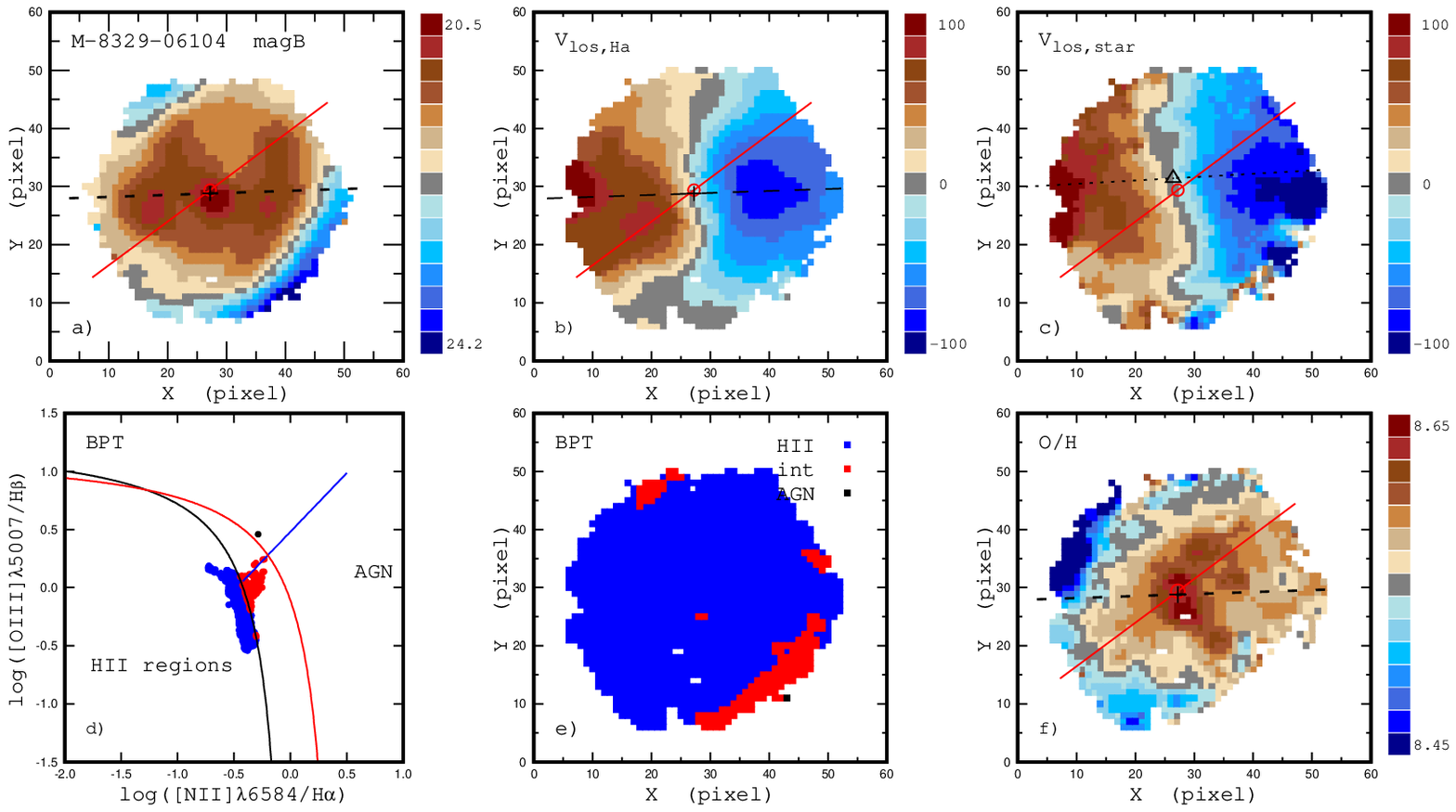}}
\caption{
Inferred properties of the MaNGA galaxy M-8329-06104.
}
\label{figure:m-8329-06104}
\end{figure}

%====================================    Fig  No  M-8450-06102 
\begin{figure}
\resizebox{1.00\hsize}{!}{\includegraphics[angle=000]{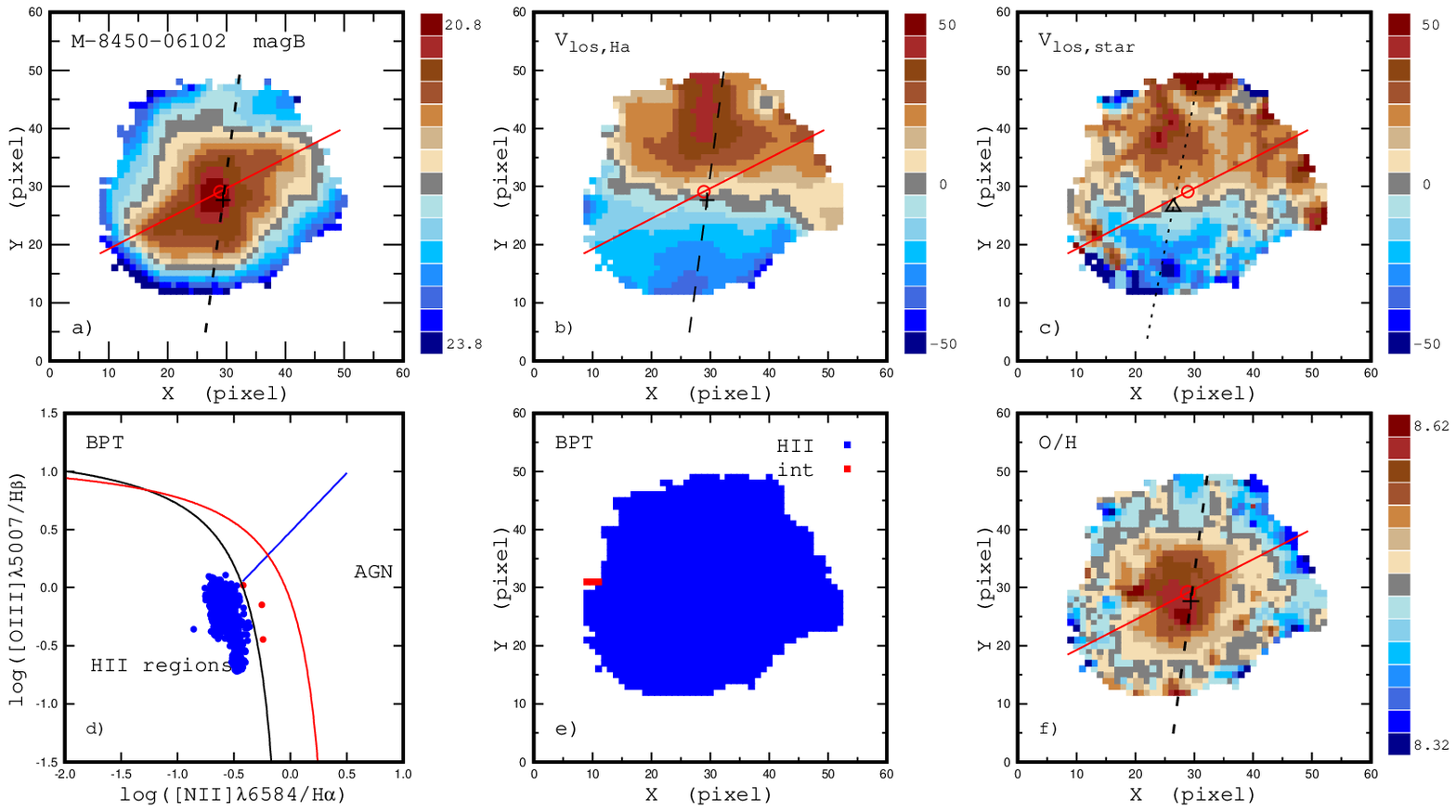}}
\caption{
Inferred properties of the MaNGA galaxy M-8450-06102.
}
\label{figure:m-8450-06102}
\end{figure}

%====================================    Fig  No  M-8451-12703 
\begin{figure}
\resizebox{1.00\hsize}{!}{\includegraphics[angle=000]{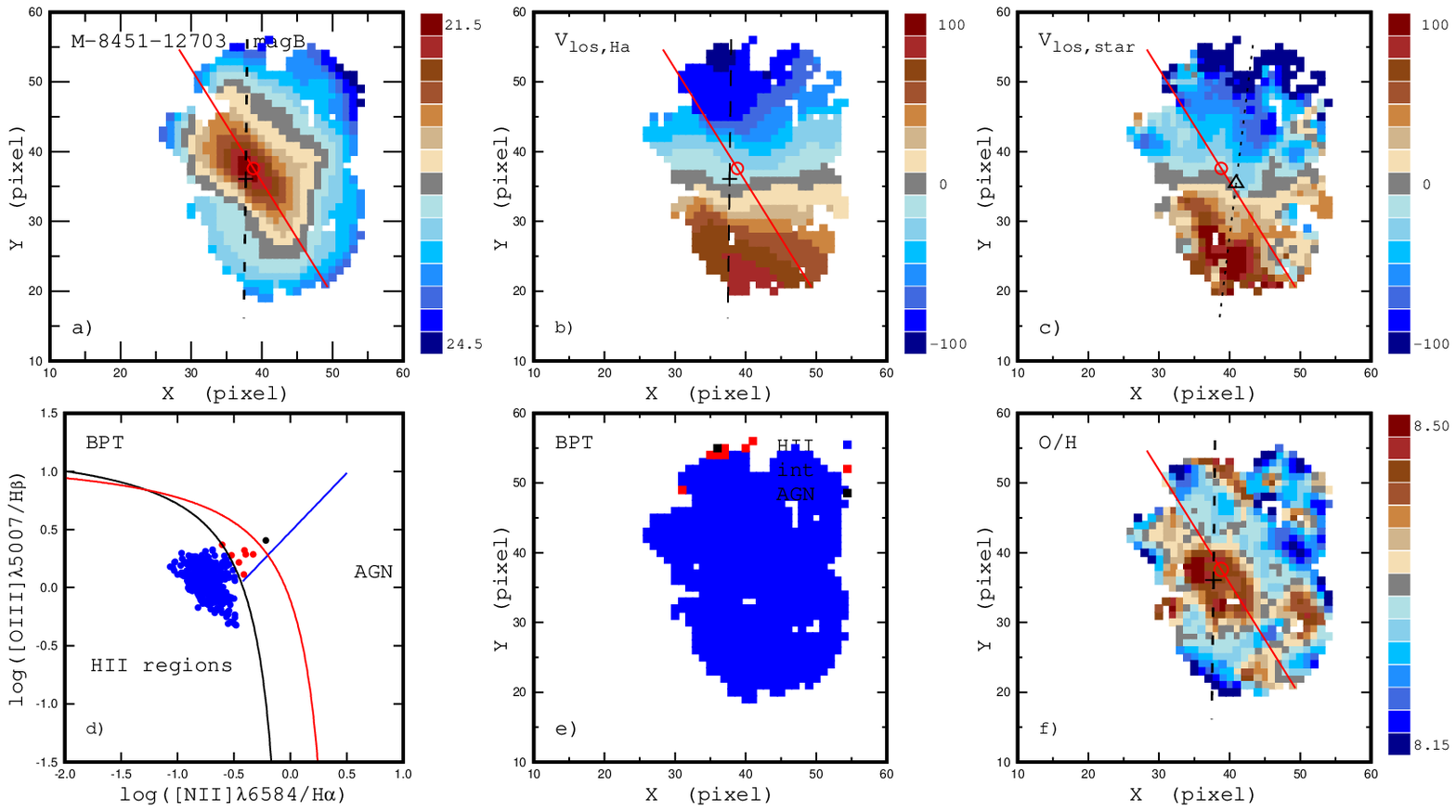}}
\caption{
Inferred properties of the MaNGA galaxy M-8451-12703.
}
\label{figure:m-8451-12703}
\end{figure}

%====================================    Fig  No  M-8483-06103 
\begin{figure}
\resizebox{1.00\hsize}{!}{\includegraphics[angle=000]{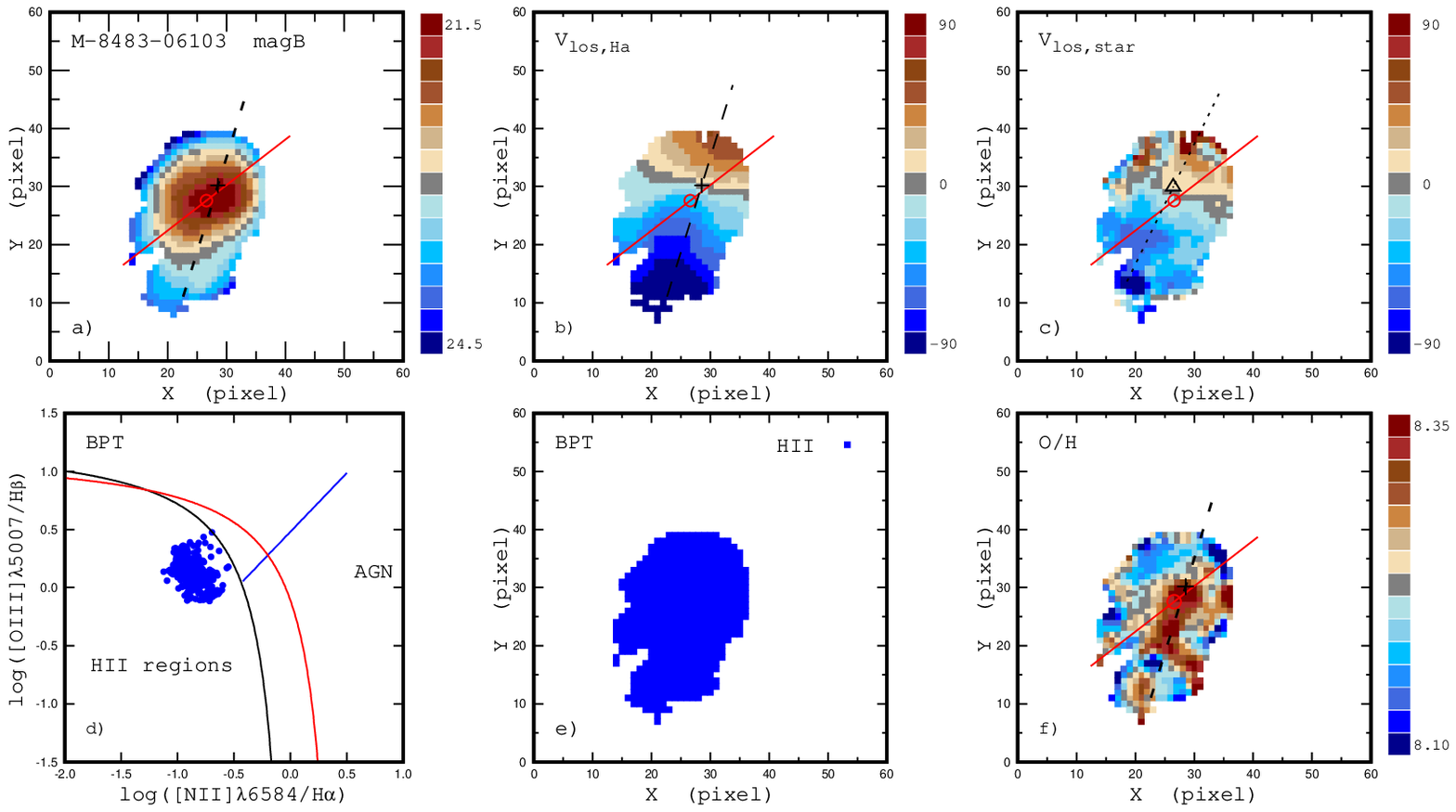}}
\caption{
Inferred properties of the MaNGA galaxy M-8483-06103. 
}
\label{figure:m-8483-06103}
\end{figure}

%====================================    Fig  No  M-8486-12702 
\begin{figure}
\resizebox{1.00\hsize}{!}{\includegraphics[angle=000]{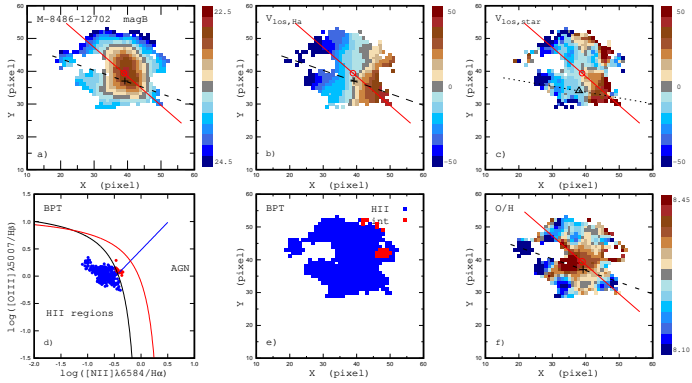}}
\caption{
Inferred properties of the MaNGA galaxy M-8486-12702. 
}
\label{figure:m-8486-12702}
\end{figure}

%====================================    Fig  No  M-8547-06102 
\begin{figure}
\resizebox{1.00\hsize}{!}{\includegraphics[angle=000]{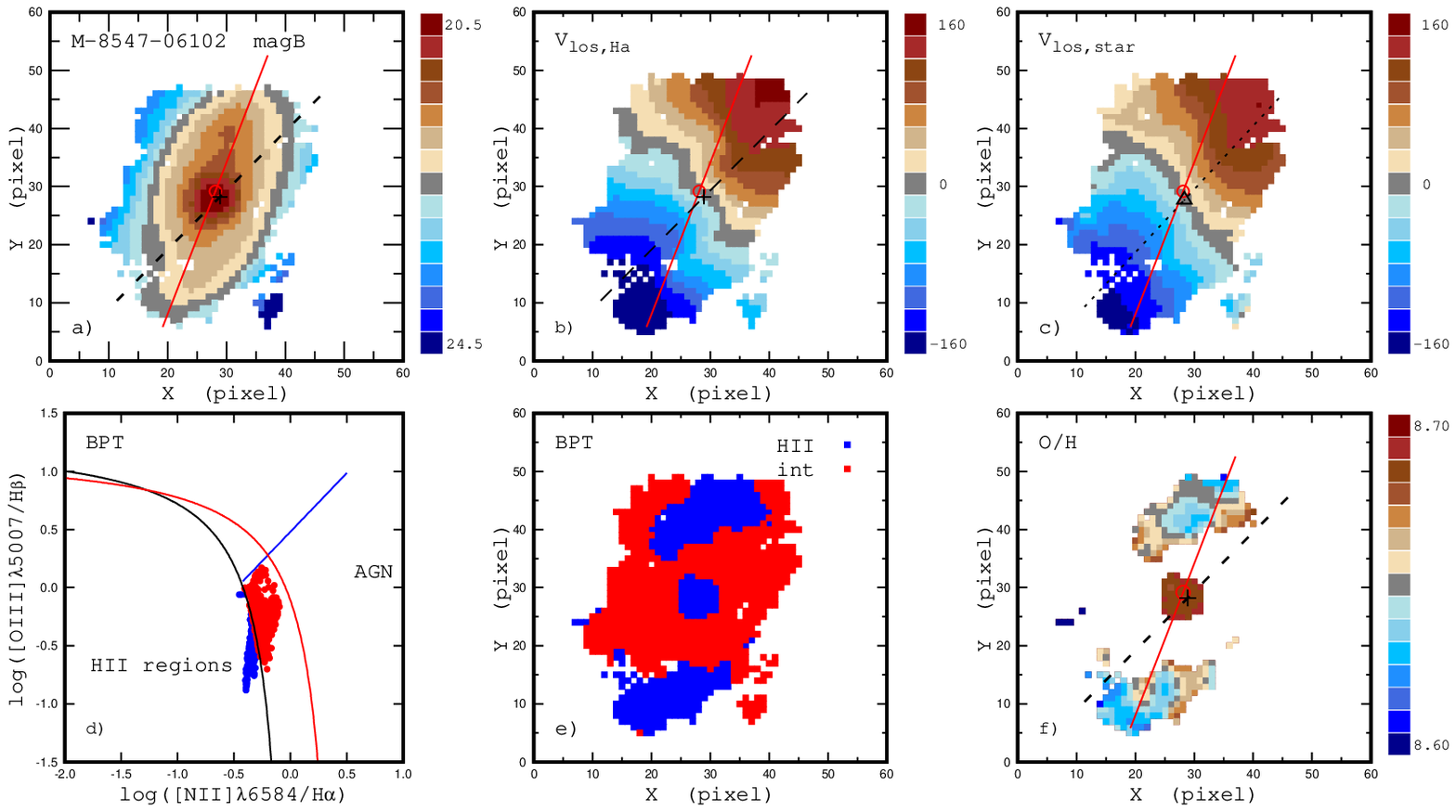}}
\caption{
Inferred properties of the MaNGA galaxy M-8547-06102. 
}
\label{figure:m-8547-06102}
\end{figure}

%====================================    Fig  No  M-8551-09102 
\begin{figure}
\resizebox{1.00\hsize}{!}{\includegraphics[angle=000]{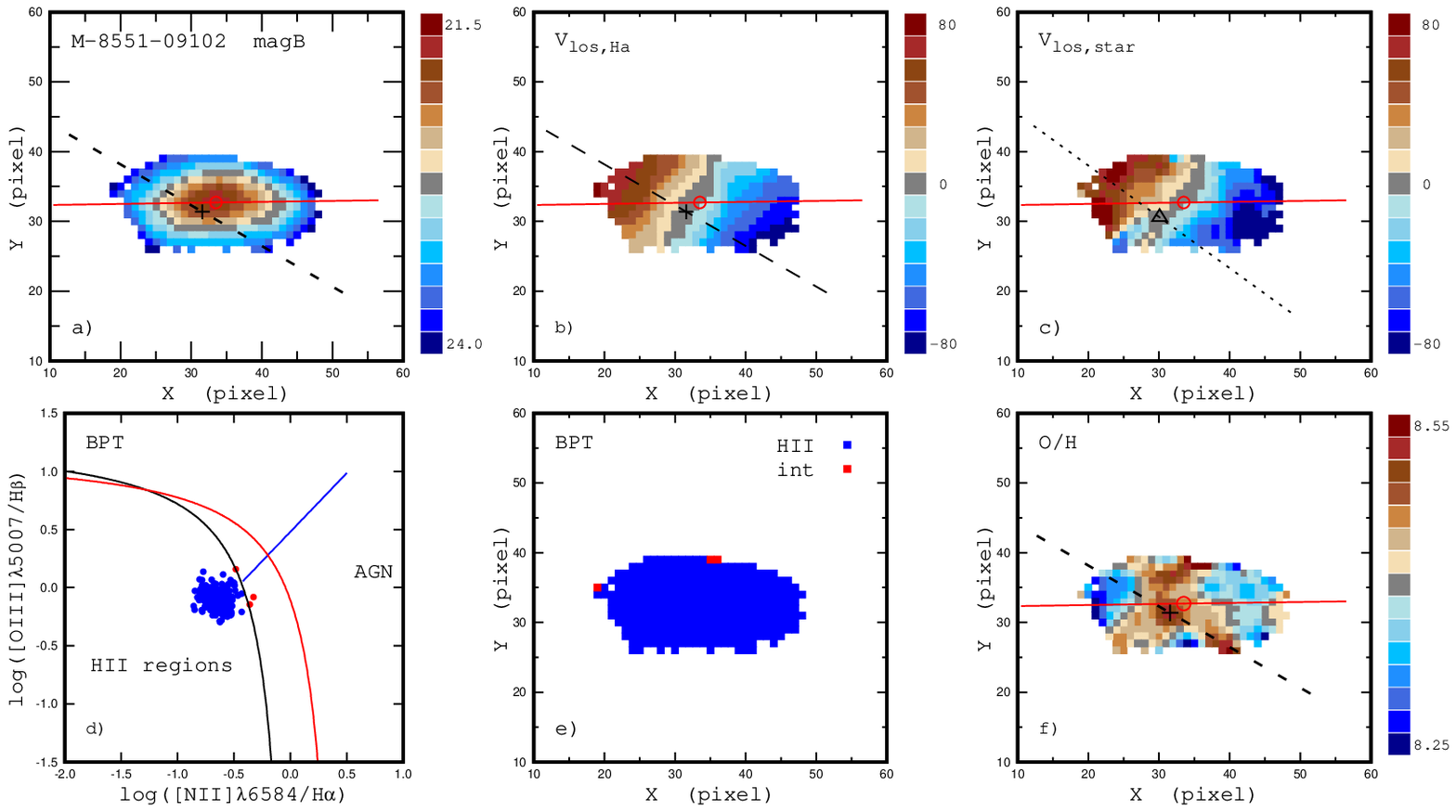}}
\caption{
Inferred properties of the MaNGA galaxy M-8551-09102. 
}
\label{figure:m-8551-09102}
\end{figure}

%====================================    Fig  No  M-8568-12703 
\begin{figure}
\resizebox{1.00\hsize}{!}{\includegraphics[angle=000]{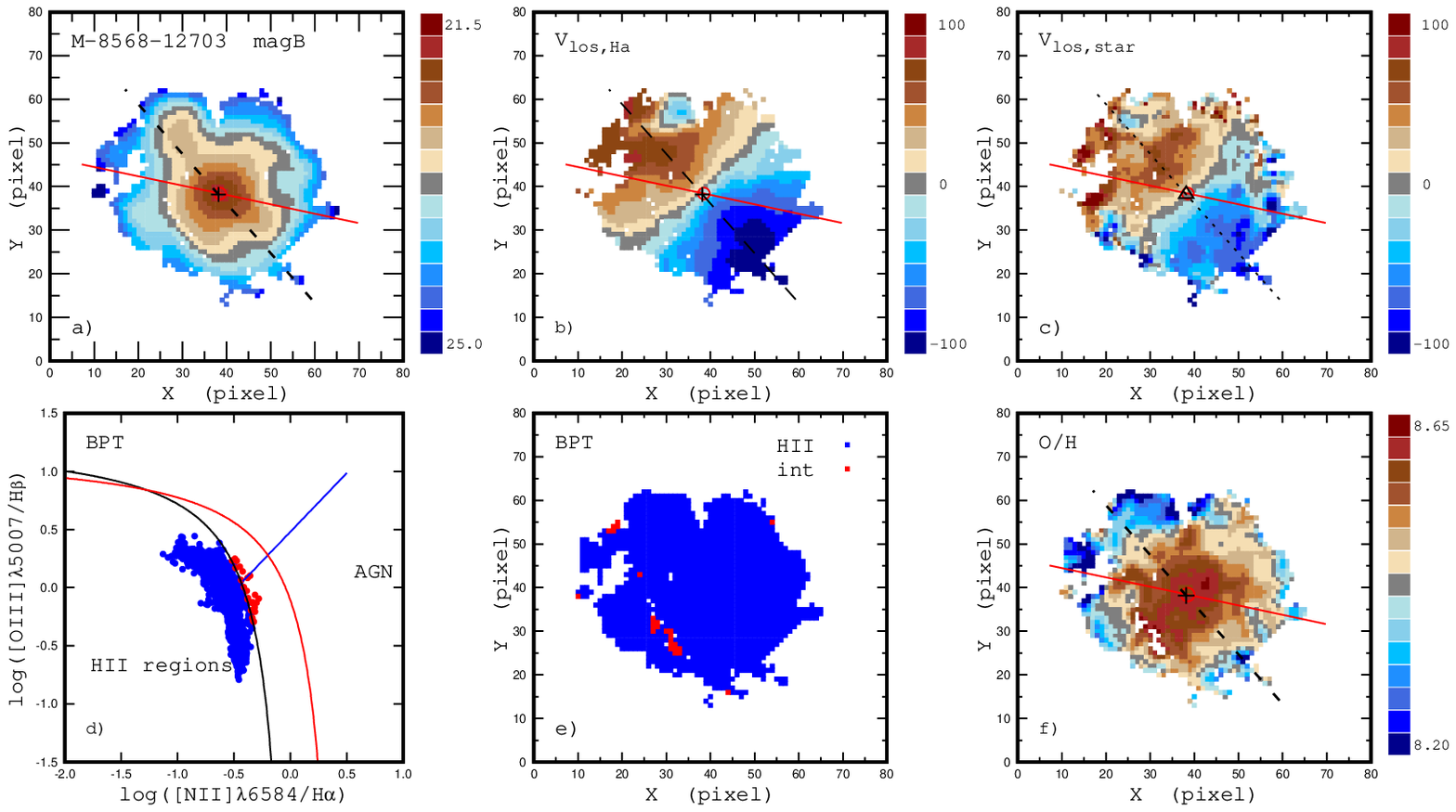}}
\caption{
Inferred properties of the MaNGA galaxy M-8568-12703. 
}
\label{figure:m-8568-12703}
\end{figure}

\end{document}